\documentclass{aa}    
\usepackage{txfonts} 
\usepackage{amssymb}
\usepackage{epstopdf}
\usepackage{epsfig}
\usepackage{url}
\usepackage{natbib}
\bibpunct{(}{)}{;}{a}{}{,} % to follow the A&A style 

\newcommand{\kms} {\mbox{${\rm km\,s^{-1}}$}}
\newcommand{\xin} {\mbox{${X_{\rm in}}~$}}
\newcommand{\xout} {\mbox{${X_{\rm out}}~$}}

\title{Deuterated water in the solar-type protostars \\
NGC~1333 IRAS~4A and IRAS~4B
\thanks{Based on observations carried out with the \textit{Herschel}/HIFI instrument, the Institut de RadioAstronomie Millim\'etrique (IRAM) 30m Telescope, the James Clerk Maxwell Telescope (JCMT), and one of the ESO telescopes at the La Silla Paranal, the Atacama Pathfinder Experiment (APEX, programme ID 090.C-0239). {\it Herschel} is an ESA space observatory with science instruments provided by
    European-led principal Investigator consortia and with important
    participation from NASA. IRAM is supported by INSU/CNRS (France), MPG (Germany), and IGN (Spain).  The JCMT is operated by the Joint Astronomy Centre on behalf of the Science and Technology Facilities Council of the United Kingdom, the Netherlands Organization for Scientific Research, and the National Research Council of Canada. APEX is a collaboration between the Max-Planck-Institut f\"ur Radioastronomie, the ESO, and the Onsala Space Observatory.
   }
}

\author{A. Coutens \inst{\ref{cesr},\ref{cesr2},\ref{nbi},\ref{museum}} \and
C. Vastel  \inst{\ref{cesr},\ref{cesr2}} \and 
S. Cabrit \inst{\ref{lerma}} \and
C. Codella  \inst{\ref{arcetri}} \and
L.\,E. Kristensen \inst{\ref{cfa}} \and
C. Ceccarelli \inst{\ref{ipag}}  \and
E.\,F. van Dishoeck \inst{\ref{leiden},\ref{garching}} \and \\
A.\,C.\,A. Boogert \inst{\ref{caltech}} \and
S. Bottinelli \inst{\ref{cesr},\ref{cesr2}} \and
A. Castets \inst{\ref{ipag}} \and
E. Caux \inst{\ref{cesr},\ref{cesr2}} \and
C. Comito \inst{\ref{koln},\ref{bonn}} \and
K. Demyk \inst{\ref{cesr},\ref{cesr2}}  \and
F. Herpin \inst{\ref{bordeaux1},\ref{bordeaux2}} \and 
B. Lefloch \inst{\ref{ipag}} \and \\
C. McCoey \inst{\ref{waterloo}} \and
J.\,C. Mottram \inst{\ref{leiden}} \and
B. Parise  \inst{\ref{bonn}} \and 
V. Taquet \inst{\ref{nasa}} \and
F.\,F.\,S. van der Tak \inst{\ref{sron},\ref{sron2}} \and
R. Visser  \inst{\ref{michigan}} \and
U.\,A. Y{\i}ld{\i}z \inst{\ref{leiden}}
}

\institute{
 Universit\'{e} de Toulouse, UPS-OMP, IRAP, Toulouse, France \\
 \email{acoutens@nbi.dk} 
\label{cesr}
\and CNRS, IRAP, 9 Av. Colonel Roche, BP 44346, F-31028 Toulouse Cedex 4, France
\label{cesr2}
\and Niels Bohr Institute, University of Copenhagen, Juliane Maries Vej 30, DK-2100 Copenhagen \O ., Denmark  
\label{nbi}
\and Centre for Star and Planet Formation, Natural History Museum of Denmark, University of Copenhagen, \O ster Voldgade 5Ð7, DK-1350 Copenhagen K., Denmark
\label{museum}
\and LERMA, Observatoire de Paris, UMR 8112 CNRS/INSU, 61 Av. de l'Observatoire, 75014 Paris, France
\label{lerma}
\and INAF - Osservatorio Astrofisico di Arcetri, Largo E. Fermi 5, 50125 Firenze, Italy
\label{arcetri}
\and Harvard-Smithsonian Center for Astrophysics, 60 Garden Street, Cambridge, MA 02138, USA
\label{cfa}
\and Institut de Plan\'{e}tologie et d'Astrophysique de Grenoble (IPAG), UMR 5274, UJF-Grenoble 1/CNRS, F-38041 Grenoble, France
\label{ipag}
\and Leiden Observatory, Leiden University, PO Box 9513, 2300 RA Leiden, The Netherlands
\label{leiden}
\and Max-Planck-Institut f\" ur Extraterrestrische Physik, Giessenbachstrasse 1, 85748 Garching, Germany
\label{garching}
\and California Institute of Technology, Infrared Processing and Analysis Center, Mail Code 100-22, Pasadena, CA 91125, USA
\label{caltech}
\and Universit\'e de Bordeaux, Laboratoire d'Astrophysique de Bordeaux, 33000 Bordeaux, France
\label{bordeaux1}
\and CNRS/INSU, UMR 5804, BP 89, 33271 Floirac Cedex, France
\label{bordeaux2}
\and Physikalisches Institut, Universit\"at zu K\" oln, Z\"ulpicher Str. 77, 50937 K\"oln, Germany
\label{koln}
\and Max-Planck-Institut f\"{u}r Radioastronomie, Auf dem H\"{u}gel 69, 53121 Bonn, Germany
\label{bonn}
\and University of Waterloo, Department of Physics and Astronomy, Waterloo, Ontario, Canada
\label{waterloo}
\and NASA Postdoctoral Program Fellow, NASA Goddard Space Flight Center, 8800 Greenbelt Road, Greenbelt, MD 20770, USA
\label{nasa}
\and SRON Netherlands Institute for Space Research, Landleven 12, 9747 AD Groningen, The Netherlands
\label{sron}
\and Kapteyn Astronomical Institute, University of Groningen, The Netherlands
\label{sron2}
\and Department of Astronomy, University of Michigan, 500 Church Street, Ann Arbor, MI 48109-1042, USA
\label{michigan}
}

\date{Received xxx / Accepted xxx}              % Activate to display a given date or no date

\abstract {
The measurement of the water deuterium fractionation is a relevant tool for understanding mechanisms of water formation and evolution from the prestellar phase to the formation of planets and comets.}
{The aim of this paper is to study deuterated water in the solar-type protostars NGC~1333 IRAS~4A and IRAS~4B, to compare their HDO abundance distributions with other star-forming regions, and to constrain their HDO/H$_2$O abundance ratios. }
{Using the \textit{Herschel}/HIFI instrument as well as ground-based telescopes, we observed several HDO lines covering a large
excitation range ($E_{\rm up}/k$ = 22--168 K) towards these protostars and an outflow position. 
Non-local thermal equilibrium radiative transfer codes were then used to determine the HDO abundance profiles in these sources.}
{
The HDO fundamental line profiles show a very broad component, tracing the molecular outflows, in addition to a narrower emission component and a narrow absorbing component. In the protostellar envelope of NGC~1333 IRAS~4A, the HDO inner (T $\geq$ 100 K) and outer (T $<$ 100 K) abundances with respect to H$_2$ are estimated with a 3$\sigma$ uncertainty at 7.5$^{+3.5}_{-3.0}$ $\times$ 10$^{-9}$ and 1.2$^{+0.4}_{-0.4}$ $\times$ 10$^{-11}$, respectively, whereas in NGC~1333 IRAS~4B they are 1$^{+1.8}_{-0.9}$ $\times$ 10$^{-8}$ and 1.2$^{+0.6}_{-0.4}$ $\times$ 10$^{-10}$, respectively. Similarly to the low-mass protostar IRAS~16293-2422, an absorbing outer layer with an enhanced abundance of deuterated water 
is required to reproduce the absorbing components seen in the fundamental lines at 465 and 894 GHz in both sources. This water-rich layer is probably extended enough to encompass the two sources, as well as parts of the outflows.
In the outflows emanating from NGC~1333 IRAS~4A, the HDO column density is estimated at about (2--4) $\times$ 10$^{13}$ cm$^{-2}$, leading to an abundance of about (0.7--1.9)~$\times$~10$^{-9}$. An HDO/H$_2$O ratio between 7 $\times$ 10$^{-4}$ and 9 $\times$ 10$^{-2}$ is also derived in the outflows. 
In the warm inner regions of these two sources, we estimate the HDO/H$_2$O ratios at about 1~$\times$~10$^{-4}$ -- 4~$\times$~10$^{-3}$. This ratio seems higher (a few \%) in the cold envelope of IRAS~4A, whose possible origin is discussed in relation to formation processes of HDO and H$_2$O.
}
{In low-mass protostars, the HDO outer abundances range in a small interval, between $\sim$10$^{-11}$ and a few 10$^{-10}$. No clear trends are found between the HDO abundance and various source parameters ($L_{\rm bol}$, $L_{\rm smm}$, $L_{\rm smm}$/$L_{\rm bol}$, $T_{\rm bol}$, $L_{\rm bol}$$^{0.6}$/$M_{\rm env}$). A tentative correlation is observed, however, between the ratio of the inner and outer abundances with the submillimeter luminosity. }


\begin{document}
\maketitle

\abstract

\section{Introduction}

The study of the deuterated isotopologues of water, HDO and D$_2$O, is helpful in order to understand water chemistry in the interstellar medium.
Indeed, the HDO/H$_2$O\footnote{The HDO/H$_2$O ratio refers to twice the water D/H abundance ratio.} ratio can be used to constrain the water formation conditions. Water can be formed by different mechanisms, both in the gas phase and on grain surfaces.
In diffuse clouds, gas-phase ion-molecule reactions lead to the formation of the H$_3$O$^+$ ion, which can dissociatively recombine to form H$_2$O \citep[e.g.,][]{Dalgarno1980,Jensen2000}. Water can also form in the gas phase through the endothermic reaction of O with H$_2$ to form OH, followed by the reaction between OH and H$_2$ \citep[e.g.,][]{Wagner1987}. Because of the endothermicity of the first reaction ($\sim$ 2000 K) and the high energy barrier of the second reaction ($\sim$ 2100 K), this mechanism is only important in regions with high temperatures ($>$ 230 K), such as hot cores and shocks \citep{Ceccarelli1996,Hollenbach1979}.
In cold and dense regions, water is mainly formed by grain-surface chemistry, through hydrogenation of atomic and molecular oxygen accreted on the grains \citep[e.g.,][]{Tielens1982,Ioppolo2008,Miyauchi2008,Dulieu2010,Cazaux2010}. Water can then be thermally desorbed when the temperature is higher than $\sim$100 K \citep{Fraser2001}, for example in the inner parts of the low-mass protostellar envelopes. It can also be released into the gas phase by non-thermal desorption mechanisms such as mechanical erosion (sputtering) in shocks \citep[e.g.,][]{Flowers1994}, or photodesorption by the interstellar radiation field or the cosmic ray induced UV-field \citep[e.g.,][]{Oberg2009,Caselli2012}.
Deuterated water is expected to be formed in a similar way, but because of deuterium fractionation effects \citep[e.g.,][]{Phillips2003}, the HDO/H$_2$O ratio is dependent on the temperature at which the water formation takes place. This ratio should then be high if water forms at low temperatures, whereas it should be low if water forms at high temperatures.

In the past decade, many attempts have been made to explain the origin of Earth's water. Its formation may be endogenous or exogenous: water adsorbed on dry silicate grains in the protosolar nebula \citep{Stimpfl2004}, delivery through asteroids, comets, planetary embryos, and planetesimals \citep{Morbidelli2000,OBrien2006,Raymond2004,Raymond2006,Raymond2009,Lunine2003,Drake2006}, and water production through oxidation of a hydrogen rich atmosphere \citep{Ikoma2006}. These sources can be investigated by studying the water D/H ratio. For example, the value toward eight Oort Cloud comets is, on average, twice that of the standard mean ocean water (SMOW, 1.56 $\times$ 10$^{-4}$) and 12 times the value of the D/H ratio in the early solar nebula ($\sim$ 2.5 $\times$ 10$^{-5}$; \citealt{Niemann1996}, \citealt{Geiss1998}). This difference led to the conclusion that comets were not the main source of the delivery of water to Earth \citep[e.g.,][]{Bockelee1998,Meier1998,Villanueva2009}, although the original value of the D/H ratio of the Earth's water is unknown, and it is unclear how that value changed during the geophysical and geochemical evolution of the Earth \citep{Campins2004,Genda2008}. Recently, the water D/H ratio was measured with the HIFI spectrometer \citep[Heterodyne Instrument for Far Infrared;][]{deGraauw2010} onboard the \textit{Herschel Space Observatory} \citep{Pilbratt2010} to 1.61 $\times$ 10$^{-4}$ in the Jupiter-family comet 103P/Hartley2 \citep{Hartogh2011}, close to the isotopic ratio measured in the Earth's oceans \citep[1.5 $\times$ 10$^{-4}$;][]{Lecuyer1998}. The determination of the HDO/H$_2$O ratio at different stages of the star formation process is then a way to determine how water evolves from the cold prestellar phase to the formation of planets and comets.

Until now, the HDO/H$_2$O ratio has been determined in four Class 0 protostars, corresponding to the main accretion phase.
At this stage, the results are quite disparate. For example, in the warm inner regions ($T$ $>$ 100 K) of IRAS~16293-2422, \citet{Parise2005} and \citet{Coutens2012,Coutens2013} estimated a warm HDO/H$_2$O ratio\footnote{We call warm HDO/H$_2$O ratios the HDO/H$_2$O ratios measured in the warm gas of the protostellar envelope (T $\geq$ 100 K). The cold HDO/H$_2$O ratios refer then to the measure in the cold gas of the outer envelope (T $<$ 100 K).}, at about a few percent, using single-dish observations, whereas \citet{Persson2013} found a much lower estimate ($\sim$ 9 $\times$ 10$^{-4}$) using interferometric data.
This source is not the only one to present divergent results that depend on the studies. Indeed, the HDO/H$_2$O ratio in the warm inner envelope of NGC~1333 IRAS~2A, first estimated by \citet{Liu2011} at $\geq$ 1\%, was determined at about 1 $\times$  10$^{-3}$ after revision of the H$_2$O abundance by \citet{Visser2013}. This result remains different from another estimate, (0.3--8) $\times$ 10$^{-2}$, by \citet{Taquet2013}.
In NGC~1333 IRAS~4A (hereafter IRAS~4A) and NGC~1333 IRAS~4B (hereafter IRAS~4B), fewer studies have been carried out and are only based on interferometric observations. In IRAS~4A, \citet{Taquet2013} found a ratio of 5 $\times$ 10$^{-3}$ -- 3 $\times$ 10$^{-2}$, and in IRAS~4B, \citet{Jorgensen2010} derived an upper limit of $\sim$ 6 $\times$ 10$^{-4}$.
For more details on the results of the different studies of HDO/H$_2$O ratios in Class 0 sources and their type of analysis, we refer the reader to Table \ref{comp_ratios} and Sect. \ref{sect_ratio}.
Singly deuterated water has also been studied in ices. Only upper limits between 5 $\times$ 10$^{-3}$ and 2 $\times$ 10$^{-2}$ have been determined \citep{Parise2003,Dartois2003}, which do not allow us to conclude on rather high ($\sim$ 10$^{-2}$) or low ($\lesssim$ 10$^{-3}$) HDO/H$_2$O ratios in grain mantles. 

Even if it is not possible to disentangle the emission from the hot corino (defined as the warm inner part of the envelope in which complex organic species are detected; \citealt{Bottinelli2004b}) and the outer envelope using single-dish telescopes, these observations can be extremely helpful to constrain the HDO/H$_2$O ratio in the outer part of the protostellar envelopes ($T$ $<$ 100 K). Indeed, the extended emission coming from the cold envelope cannot be probed with interferometers. Only two Class~0 sources were consequently studied for their cold HDO/H$_2$O ratios$^2$. It was determined to be between 3 $\times$ 10$^{-3}$ and 1.5 $\times$ 10$^{-2}$ in IRAS~16293-2422 by \citet{Coutens2012,Coutens2013}, and between 9 $\times$ 10$^{-3}$ and 1.8 $\times$ 10$^{-1}$ in NGC~1333 IRAS~2A by \citet{Liu2011}.
The fundamental HDO 1$_{1,1}$--0$_{0,0}$ transition at 894 GHz observed with \textit{Herschel}/HIFI at high sensitivity provides particularly strong constraints on the outer HDO abundance.
In IRAS~16293-2422, this line shows a specific line profile with a very deep absorption in addition to emission \citep{Coutens2012}.
Combined with the observation of the other fundamental HDO 1$_{0,1}$--0$_{0,0}$ transition at 465 GHz with the JCMT, \citet{Coutens2012} showed that a cold water-rich absorbing layer surrounds this source. Without this layer, the deep absorbing components cannot be reproduced. Similar conclusions were obtained for the D$_2$O absorbing lines observed with \textit{Herschel} \citep{Vastel2010,Coutens2013}. However, the origin of this absorbing layer is not clearly determined. It has been suggested, for example, that it could result from an equilibrium between photodesorption and photodissociation by the UV interstellar radiation field, as predicted by \citet{Hollenbach2009}. 
It would thus be helpful to know if this layer is observed around other protostars. 
Indeed, the ubiquity of this layer would give clues to the nature of this layer.

A collaboration between three \textit{Herschel} Key Programs: CHESS \citep[Chemical HErschel Surveys of Star forming regions;][]{Ceccarelli2010}, WISH \citep[Water In Star-forming regions with Herschel;][]{VanDishoeck2011}, and HEXOS \citep[\textit{Herschel}/HIFI observations of EXtraOrdinary Sources: The Orion and Sagittarius B2 Star-forming Regions;][]{Bergin2010} was set up to investigate the water chemistry in star-forming regions.
As part of this collaboration, new observations of HDO were carried out towards two low-mass protostars, IRAS~4A and IRAS~4B, allowing us to estimate the HDO abundance profiles in these sources. These results are useful to derive, in particular, the HDO/H$_2$O ratio in the outer envelope of these protostars.
These two sources are located in the NGC~1333 reflection nebula in the Perseus molecular cloud. They are separated by 31$\arcsec$ ($\sim$ 7500 AU), and 
IRAS~4A is a binary system with a separation of $\sim$1.8$\arcsec$ between its two components \citep{Lay1995,Jorgensen2007}.
These protostars are well known for the complex organic chemistry observed in their hot corinos \citep{Bottinelli2004b,Bottinelli2007,Sakai2006,Persson2012} and for their outflows detected in particular in CO, SiO, and CS \citep[see, e.g.,][]{Blake1995,Lefloch1998,Choi2005,Yildiz2012} and in H$_2$O \citep{Kristensen2010,Kristensen2012}. The distance to the NGC~1333 region is uncertain (see \citealt{Curtis2010} for more details). We adopt here the distance of 235$\pm$18 pc determined by \citet{Hirota2008} using very-long-baseline interferometry (VLBI) parallax measurements of water masers in SVS 13 located in the same cloud. 

The paper is organized as follows. First, we present the observations in Sect. 2. Then we describe the modeling and show the results in Sect. 3. Finally, we discuss the results in Sect. 4 and conclude in Sect. 5.

\section{Observations}

\begin{table*}[!ht]
\begin{center}
\caption{Parameters for the observed HDO lines$^{(1)}$.} 
\label{obs}
\begin{tabular}{c c c c c c c c c c}
\hline
\hline	
 Frequency & $J_{\rm Ka,Kc}$  & $E_{\rm up}/k$ & $A_{\rm ij}$ &  Telescope & Beam & $F_{\rm eff}$ & $B_{\rm eff}$ & rms$^{(2)}$ & $\int T_{\rm mb}$ d$\varv$$^{(3)}$  \\
(GHz) & & (K)  & (s$ ^{-1}$) &  & size (\arcsec) & & & (mK) & (K km s$^{-1}$) \\
\hline
\multicolumn{10}{c}{NGC~1333 IRAS~4A} \\
\hline
  80.5783 & 1$_{1,0}$--1$_{1,1}$ & 47 & 1.32 $\times$ 10$^{-6}$ & IRAM-30m & 31.2 & 0.95 & 0.78 & 4 & 0.025 $\pm$ 0.005   \\    
 225.8967 & 3$_{1,2}$--2$_{2,1}$ & 168 & 1.32 $\times$ 10$^{-5}$ & IRAM-30m & 11.1 & 0.91 & 0.54 & 9 & 0.26 $\pm$ 0.03   \\    
 241.5616 & 2$_{1,1}$--2$_{1,2}$ & 95 & 1.19 $\times$ 10$^{-5}$ & IRAM-30m & 10.4 & 0.90 & 0.57 & 18 &  0.25 $\pm$ 0.04  \\    
 464.9245 & 1$_{0,1}$--0$_{0,0}$ & 22 & 1.69 $\times$ 10$^{-4}$ & JCMT & 10.8 & -- & 0.44$^{(4)}$ & 53 &  1.8 $\pm$ 0.2$^{(5)}$ \\    
 599.9267 & 2$_{1,1}$--2$_{0,2}$ & 95 & 3.45 $\times$ 10$^{-3}$ & HIFI 1b & 35.9 & 0.96 & 0.75 & 7 &  $\leq$ 0.05 \\   
 893.6387 & 1$_{1,1}$--0$_{0,0}$ & 43 & 8.35 $\times$ 10$^{-3}$ & HIFI 3b & 24.1 & 0.96 & 0.74 & 4 &  0.55 $\pm$ 0.02$^{(5)}$   \\  
\hline
\multicolumn{10}{c}{outflow position of NGC~1333 IRAS~4A} \\
\hline
 599.9267 & 2$_{1,1}$--2$_{0,2}$ & 95 & 3.45 $\times$ 10$^{-3}$ & HIFI 1b & 35.9 & 0.96 & 0.75 & 7 & $\leq$ 0.05 \\   
 893.6387 & 1$_{1,1}$--0$_{0,0}$ & 43 & 8.35 $\times$ 10$^{-3}$ & HIFI 3b & 24.1 & 0.96 & 0.74 & 4 & 0.27 $\pm$ 0.01$^{(5)}$ \\  
\hline
\multicolumn{10}{c}{NGC~1333 IRAS~4B} \\
\hline
 80.5783 & 1$_{1,0}$--1$_{1,1}$ & 47 & 1.32 $\times$ 10$^{-6}$ & IRAM-30m & 31.2 & 0.95 & 0.81   & 4 &  0.031 $\pm$ 0.009 \\    
 225.8967 & 3$_{1,2}$--2$_{2,1}$ & 168 & 1.32 $\times$ 10$^{-5}$ & IRAM-30m & 11.1 & 0.91 & 0.61 & 9 &  $\leq$ 0.05  \\    
 241.5616 & 2$_{1,1}$--2$_{1,2}$ & 95 & 1.19 $\times$ 10$^{-5}$ & IRAM-30m & 10.4 & 0.90 & 0.57 & 4 & 0.051 $\pm$ 0.008 \\    
 464.9245 & 1$_{0,1}$--0$_{0,0}$ & 22 & 1.69 $\times$ 10$^{-4}$ & APEX & 13.4  & 0.95 & 0.60 & 40 & 1.2 $\pm$ 0.1$^{(5)}$  \\   
 599.9267 & 2$_{1,1}$--2$_{0,2}$ & 95 & 3.45 $\times$ 10$^{-3}$ & HIFI 1b & 35.9 & 0.96 & 0.75 & 6 & 0.07 $\pm$ 0.01 \\   
 893.6387 & 1$_{1,1}$--0$_{0,0}$ & 43 & 8.35 $\times$ 10$^{-3}$ & HIFI 3b & 24.1 & 0.96 & 0.74 & 4  &  0.40 $\pm$ 0.01$^{(5)}$ \\  
\hline
\end{tabular}
\end{center}
$^{(1)}$ The frequencies, the upper state energies ($E_{\rm up}$), and the Einstein coefficients ($A_{\rm ij}$) of HDO come from the spectroscopic catalog JPL \citep{Pickett1998}. \\
$^{(2)}$ The rms (in $T_{\rm mb}$) is computed for a spectral resolution of 0.5\,km\,s$^{-1}$. \\
$^{(3)}$ The calibration uncertainties are not taken into account in the flux uncertainties.\\
$^{(4)}$ This value corresponds to the ratio between the beam efficiency and the forward efficiency.\\
$^{(5)}$ For these lines, the integrated intensity is directly estimated, combining both emission and absorption components. To derive the flux of the different components (broad and narrow emission components and absorption component), we refer to Table \ref{decomp}. 
\end{table*}%

\begin{table*}[!ht]
\begin{center}
\caption{HDO line components observed towards IRAS~4A and IRAS~4B$^{(1)}$.} 
\label{decomp}
\begin{tabular}{ @{~}l@{~~}c | c@{~~~}c@{~~~}c | c@{~~~}c@{~~~}c | c@{~~~}c@{~~~}c@{~}}
\hline
\hline
\multicolumn{2}{c}{ } & \multicolumn{3}{|c}{Broad emission component} & \multicolumn{3}{|c}{Narrow emission component} & \multicolumn{3}{|c}{Absorption component}\\
\hline
 Frequency & $J_{\rm Ka,Kc}$  & $T_{\rm mb}^{\rm peak}$ & $FWHM$ & $\varv_{\rm LSR}$ & $T_{\rm mb}^{\rm peak}$ & $FWHM$ & $\varv_{\rm LSR}$ & $T_{\rm mb}^{\rm peak}$ & $FWHM$ & $\varv_{\rm LSR}$ \\
(GHz) & & (K)  & (km s$^{-1}$) &(km s$^{-1}$) & (K)  & (km s$^{-1}$) &(km s$^{-1}$) & (K)  & (km s$^{-1}$) &(km s$^{-1}$)\\
\hline
\multicolumn{5}{c}{} & \multicolumn{3}{c}{NGC~1333 IRAS~4A} \\
\hline
80.5783 & 1$_{1,0}$-1$_{1,1}$ & -- & -- & -- & 0.013 $\pm$ 0.003 & 1.4 $\pm$ 0.4 & 5.8 $\pm$ 0.2 & -- & -- & -- \\    
225.8967 & 3$_{1,2}$--2$_{2,1}$ & -- & -- & -- & 0.032 $\pm$ 0.004 & 7.9 $\pm$ 0.9  & 6.0 $\pm$ 0.4 & -- & -- & --  \\    
241.5616 & 2$_{1,1}$--2$_{1,2}$ & -- & -- & -- & 0.047 $\pm$ 0.006 & 5.4 $\pm$ 0.9 & 7.9 $\pm$ 0.4 & -- & -- & --     \\    
 464.9245 & 1$_{0,1}$--0$_{0,0}$ &  0.12 $\pm$ 0.02 & 15.9$^{(2)}$  & 5.9$^{(2)}$  & 0.25 $\pm$ 0.05 & 1.8 $\pm$ 0.5 & 7.0 $\pm$ 0.3 & $-$0.62 $\pm$ 0.07 & 0.9 $\pm$ 0.1 & 7.7 $\pm$ 0.1 \\       
 893.6387 & 1$_{1,1}$--0$_{0,0}$ &  0.040 $\pm$ 0.002 & 15.9 $\pm$ 0.6 & 5.9 $\pm$ 0.2 & 0.14 $\pm$ 0.01 & 2.0 $\pm$ 0.1 & 7.2 $\pm$ 0.1 & $-$0.38 $\pm$ 0.01 & 1.2 $\pm$ 0.1 & 7.6 $\pm$ 0.1 \\  
\hline
\multicolumn{5}{c}{} & \multicolumn{3}{c}{NGC~1333 IRAS~4B} \\
\hline
80.5783 & 1$_{1,0}$--1$_{1,1}$ &  -- & -- & -- & 0.006 $\pm$ 0.002 & 5 $\pm$ 2 & 8.2 $\pm$ 0.7 & -- & -- & -- \\
241.5616 & 2$_{1,1}$--2$_{1,2}$ & -- & -- & -- & 0.014 $\pm$ 0.002 & 4.3 $\pm$ 0.8 & 6.9 $\pm$ 0.3 & -- & -- & --  \\
464.9245 & 1$_{0,1}$--0$_{0,0}$ & 0.08 $\pm$ 0.02 & 10.0$^{(2)}$ & 6.8$^{(2)}$ & 0.6 $\pm$ 0.4 & 1.1 $\pm$ 0.2  & 7.0 $\pm$ 0.4 &  $-$0.6 $\pm$ 0.6 & 0.7 $\pm$ 0.3 & 7.4 $\pm$ 0.1 \\       
599.9267 & 2$_{1,1}$--2$_{0,2}$ &  -- & --   & --  &  0.029 $\pm$ 0.003 & 2.7 $\pm$ 0.4 & 7.6 $\pm$ 0.2 & -- & -- & -- \\   
893.6387 & 1$_{1,1}$--0$_{0,0}$ & 0.044 $\pm$ 0.003 & 10.0 $\pm$ 0.4 & 6.8 $\pm$ 0.2 & 0.22 $\pm$ 0.03 & 2.0 $\pm$ 0.1 & 7.4 $\pm$ 0.1 & $-$0.39 $\pm$ 0.01 & 1.3 $\pm$ 0.1 & 7.6 $\pm$ 0.1 \\  
\hline
\end{tabular}
\end{center}
$^{(1)}$ Obtained from Gaussian fits to each component using the Levenberg-Marquardt algorithm \citep{Levenberg1944,Marquardt1963}. The uncertainties are statistical and do not include the calibration uncertainties. \\
$^{(2)}$ Fixed parameters. \\
\end{table*}

\begin{figure}[!ht]
\begin{center}
\includegraphics[scale=0.35]{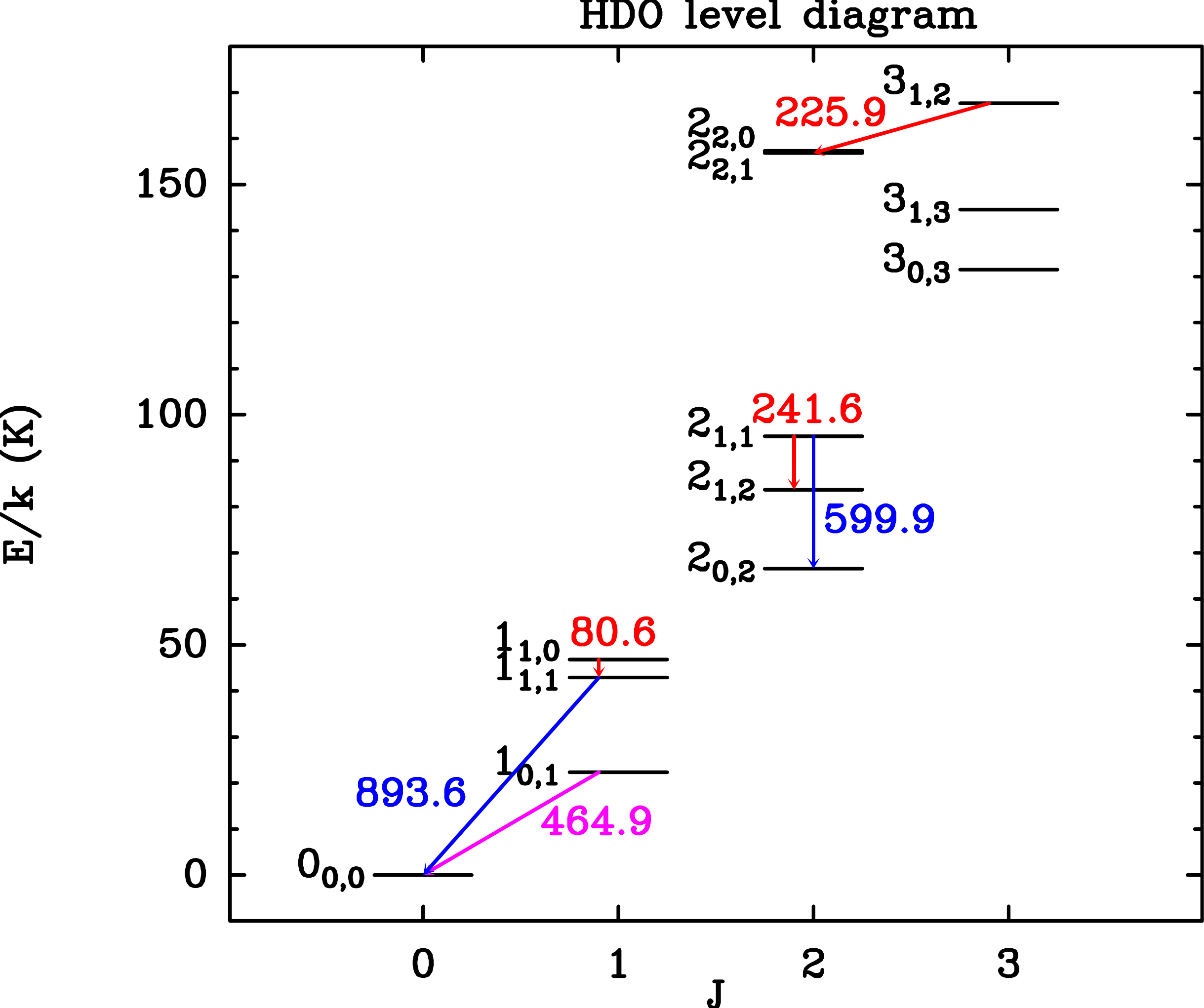}
\caption{Energy level diagram of the HDO lines. In \textit{red}, IRAM-30m observations; in \textit{magenta}, JCMT/APEX observation; in \textit{blue}, HIFI observations. The frequencies are written in GHz.}
\label{hdo_level_diagram}
\end{center}
\end{figure}

The various transitions observed towards IRAS~4A and IRAS~4B are shown in the energy level diagram in Fig. \ref{hdo_level_diagram} and summarized in Table~\ref{obs}. The original observations (without subtraction of the broad outflow component) of the 1$_{1,1}$--0$_{0,0}$ and 1$_{0,1}$--0$_{0,0}$ fundamental transitions are presented in Fig. \ref{hdo_outflow}. The other transitions are shown with the modeling in Figs. \ref{ratran_bestfit_iras4a} and \ref{ratran_bestfit_iras4b}.

\subsection{HIFI data}

In the framework of a collaboration between the CHESS, WISH, and HEXOS \textit{Herschel} Key programs, two HDO transitions were observed with the HIFI instrument towards the solar-type protostars IRAS~4A and IRAS~4B: the fundamental 1$_{1,1}$--0$_{0,0}$ line at 894 GHz with $E_{\rm up}$ = 43 K, and the 2$_{1,1}$--2$_{0,2}$ line at 600 GHz with $E_{\rm up}$ = 95 K (see Table \ref{obs}).
The targeted coordinates are $\alpha_{2000}$ = 3$^h$ 29$^m$ 10$\fs$5, $\delta_{2000}$ = 31$\degr$ 13$\arcmin$ 30.9$\arcsec$ for IRAS~4A, and $\alpha_{2000}$ = 3$^h$ 29$^m$ 12$\fs$0, $\delta_{2000}$ = 31$\degr$ 13$\arcmin$ 8.1$\arcsec$ for IRAS~4B. 
The two transitions were also observed in the red-shifted part of the outflow emanating from IRAS~4A at $\alpha_{2000}$ = 3$^h$ 29$^m$ 10$\fs$8, $\delta_{2000}$ = 31$\degr$ 13$\arcmin$ 50.9$\arcsec$. It corresponds to an offset position (+4$\arcsec$, +20$\arcsec$) with respect to IRAS~4A. This position was chosen using the map of the CO 6--5 line in Fig. 3 of \citet{Yildiz2012}, so that the telescope HPBW (half power beam width) at 600 and 894 GHz (36 and 24$\arcsec$, respectively) do not include the position of the IRAS~4A source. 

The pointed observations were obtained in August 2011, using the HIFI double beam switch (DBS) fast chop mode with optimization of the continuum. The DBS reference positions were situated at 3$\arcmin$ from the source. We checked that no line was detected in the OFF-position spectra.
The HIFI wide band spectrometer (WBS) was used, providing a spectral resolution of 1.1 MHz (0.55\,km\,s$^{-1}$ at 600 GHz and 0.37\,km\,s$^{-1}$ at 894 GHz). 
The data were processed using the standard HIFI pipeline up to frequency and amplitude calibrations (level 2) with the ESA-supported package HIPE 7.1 \citep{ott2010}. 
After being inspected separately, the H and V polarizations were averaged weighting them by the observed noise, using the GILDAS/CLASS\footnote{\url{http://www.iram.fr/IRAMFR/GILDAS/}} software. 
The forward efficiency is about 0.96 and the main beam efficiency about 0.74--0.75 \citep{Roelfsema2012}. The HIFI instrument uses double sideband receivers. A sideband gain ratio of 1 is assumed to estimate the continuum value at 894 GHz \citep{Roelfsema2012}, necessary in the modeling of the deep absorption line. 
The continuum was fitted with a polynomial of degree 1. Standing waves are not visible in the observations. 
From \citet{Roelfsema2012}, we thus estimate the uncertainties on the continuum level to be less than 10\% for both IRAS 4A and IRAS 4B observations. 
Observations were also carried out with the high resolution spectrometer (HRS). The WBS and HRS observations are in agreement (see Fig. \ref{hrs}).

\begin{figure}[!t]
\begin{center}
\includegraphics[scale=0.42]{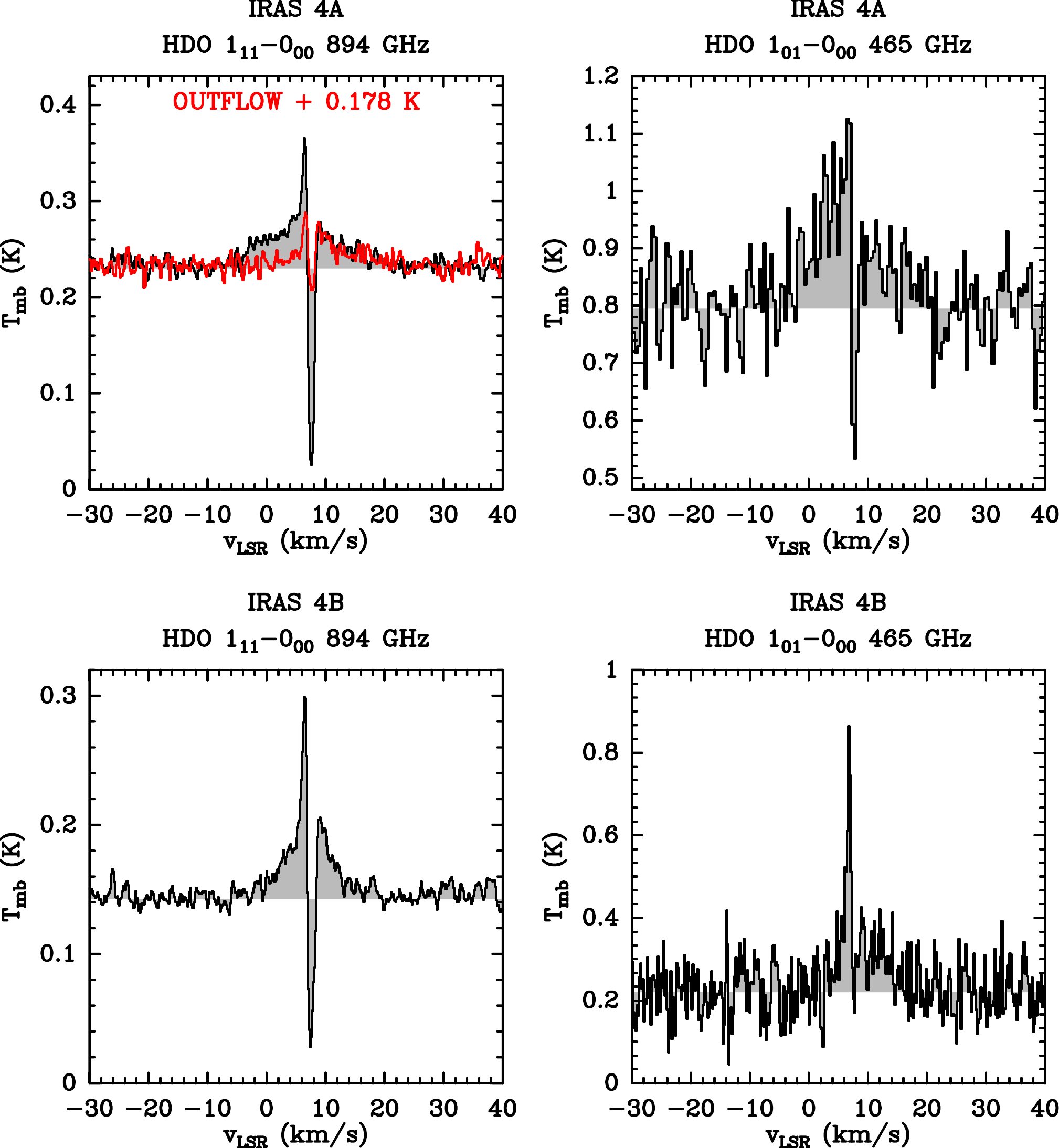}
\caption{Upper-left panel: HIFI observations of the fundamental line at 894 GHz towards the protostar IRAS~4A (black) and an outflow position (red, discussed in Sect. \ref{abs_layer}). The spectrum at the outflow position has been shifted vertically by 0.178 K. Upper-right panel: JCMT observations of the fundamental line at 465 GHz towards IRAS~4A. Lower-left panel: HIFI observations of the fundamental line at 894 GHz towards the protostar IRAS~4B. Lower-right panel: APEX observations of the fundamental line at 465 GHz towards IRAS~4B. The continuum refers to SSB data for each panel. } 
\label{hdo_outflow}
\end{center}
\end{figure}

\begin{figure}[!t]
\begin{center}
\includegraphics[scale=0.42]{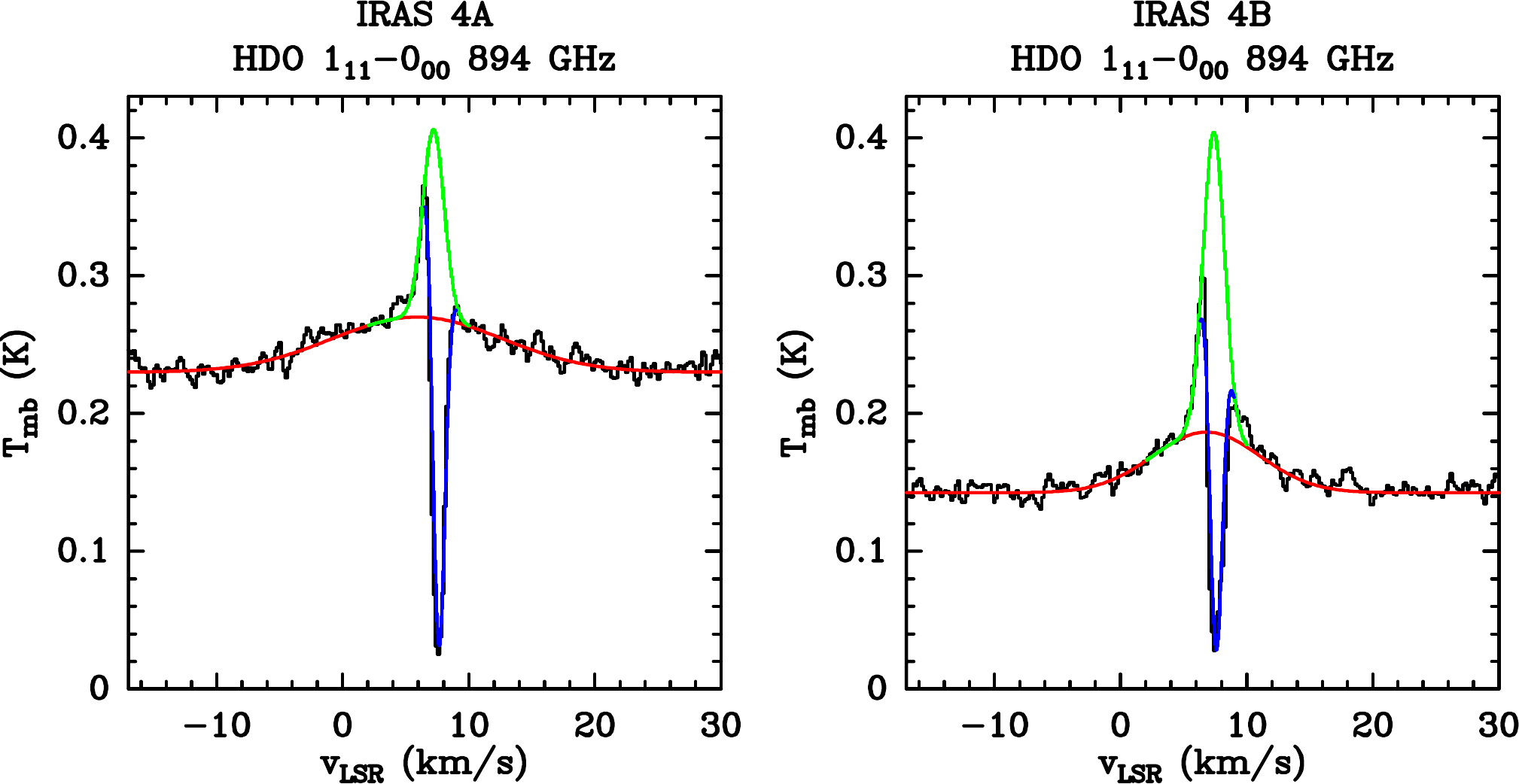}
\caption{Decomposition of the 1$_{1,1}$--0$_{0,0}$ fundamental transitions observed towards IRAS~4A and IRAS~4B with HIFI in three Gaussians: a broad emission component (red), a narrower emission component (green), and a narrow absorption component (blue). The parameters of the Gaussian fits are given in Table \ref{decomp}.} 
\label{hdo_decomp}
\end{center}
\end{figure}

The line profiles observed at 894 GHz clearly show a broad emission component tracing the outflows, a narrower emission component tracing the envelope, and a deep narrow absorption component (see Fig. \ref{hdo_decomp}). The parameters of the three components, fitted by Gaussians using the CASSIS\footnote{CASSIS (\url{http://cassis.irap.omp.eu}) has been developed by IRAP-UPS/CNRS.} software, are presented in Table \ref{decomp}. 
Fig. \ref{hdo_outflow} superposes the HIFI HDO spectra towards IRAS~4A and the position in the red outflow lobe. The spectrum in the red lobe is similar to that towards IRAS~4A except that the blue outflow wing is no longer seen. The absorption appears less deep because of the lower continuum level at the outflow position.

\subsection{JCMT data}

The HDO 1$_{0,1}$--0$_{0,0}$ fundamental transition at 465 GHz (Table~\ref{obs}) was observed towards the source IRAS~4A with the JCMT in September 2004 (project M04BN06).
The spectral resolution of the observations is 0.1 km\,s$^{-1}$. 
As in the case of the other fundamental transition at 894 GHz observed with HIFI, three components are observed: a broad component tracing the outflows, a narrower emission line, and a deep absorption component (see Fig. \ref{hdo_outflow}). The continuum level is $\sim$ 0.8 K with an uncertainty of 13\% obtained from the comparison of the spectra in the dataset.
The FWHM (full width at half maximum) and the peak-intensity velocity of the Gaussian fitted on the broad component at 894 GHz are consistent with the data at 465 GHz. 
The fit results from the higher signal-to-noise ratio profile at 894 GHz were then fixed for the 465 GHz profile to determine the intensity of the broad component (see Table \ref{decomp}).
Keeping parameters free would not affect the results significantly.

\subsection{APEX data}

The HDO 1$_{0,1}$--0$_{0,0}$ fundamental transition was also observed towards IRAS~4B with the Swedish Heterodyne Facility Instrument (SHeFI) receiver at 460 GHz of the APEX telescope. 
The observations were carried out in October 2012 using the wobbler symmetric switching mode with an amplitude of 150$\arcsec$, resulting in OFF positions at 300$\arcsec$ from the source.
The beam efficiency and the forward efficiency are shown in Table \ref{obs}. This fundamental line again shows a broad emission component, a narrow emission component, and a weak absorbing component. The absorbing component appears weaker than in IRAS~4A, because the absorbing line is slightly shifted with respect to the velocity of the narrower emission component, leading to less absorption than there would be if the velocity of the absorption was exactly the same as the velocity of the emission. 
A continuum level cannot be extracted with precision from these observations. It is estimated at about 0.2 K according to the model predictions (see Sect. \ref{sect_iras4b}).
The continuum level would then be lower in IRAS~4B than in IRAS~4A ($\sim$ 0.8 K), which also implies a shallower absorption in IRAS~4B for the same column density of the absorber.

\subsection{IRAM data}

Three additional transitions at 81 (1$_{1,0}$--1$_{1,1}$), 226 (3$_{1,2}$--2$_{2,1}$), and 242 GHz (2$_{1,1}$--2$_{1,2}$) were observed with the IRAM-30m telescope towards IRAS~4A. 
The observations were carried out in November 2004 for the 81 and 226 GHz lines with the VESPA autocorrelator in wobbler switching mode, whereas the 242 GHz transition was observed in April 2012 in position switching mode using the fast Fourier transform spectrometer (FTS) at a 200 kHz resolution.
The spectral resolution is 0.14, 0.10, and 0.24 km\,s$^{-1}$ for the 81, 226, and 242 GHz transitions respectively. For clarity, the spectra shown hereafter were smoothed to a resolution of $\sim$0.4--0.6 km\,s$^{-1}$.

These three transitions were also observed towards IRAS~4B in January 2013. The observations were carried out in position switching mode using the FTS with a fine spectral resolution of 50 kHz.
The beam efficiencies and forward efficiencies for the different observations are shown in Table \ref{obs}.
\\

Table \ref{decomp} summarizes, for each line, the Gaussian parameters of the different components derived with CASSIS. The FWHM of the narrower emission component is different depending on the line. 
Indeed these sources undergo an infall which necessarily leads to higher FWHM for the excited lines, such as the 3$_{1,2}$--2$_{2,1}$ and 2$_{1,1}$--2$_{1,2}$ transitions that probe the warm inner regions, and smaller FWHM for the fundamental lines that probe a colder medium. Moreover, the determination of the FWHM for the fundamental lines could be underestimated because of the blending between the narrow emission and absorption components. Only the broad emission component can be clearly extracted thanks to the absence of blending at high velocities ($\Delta \varv$ $>$ 3 km\,s$^{-1}$).

\section{Modeling and results}

\subsection{The protostellar envelope of NGC~1333 IRAS~4A}

 \begin{figure}[!t]
\begin{center}
\includegraphics[scale=0.33]{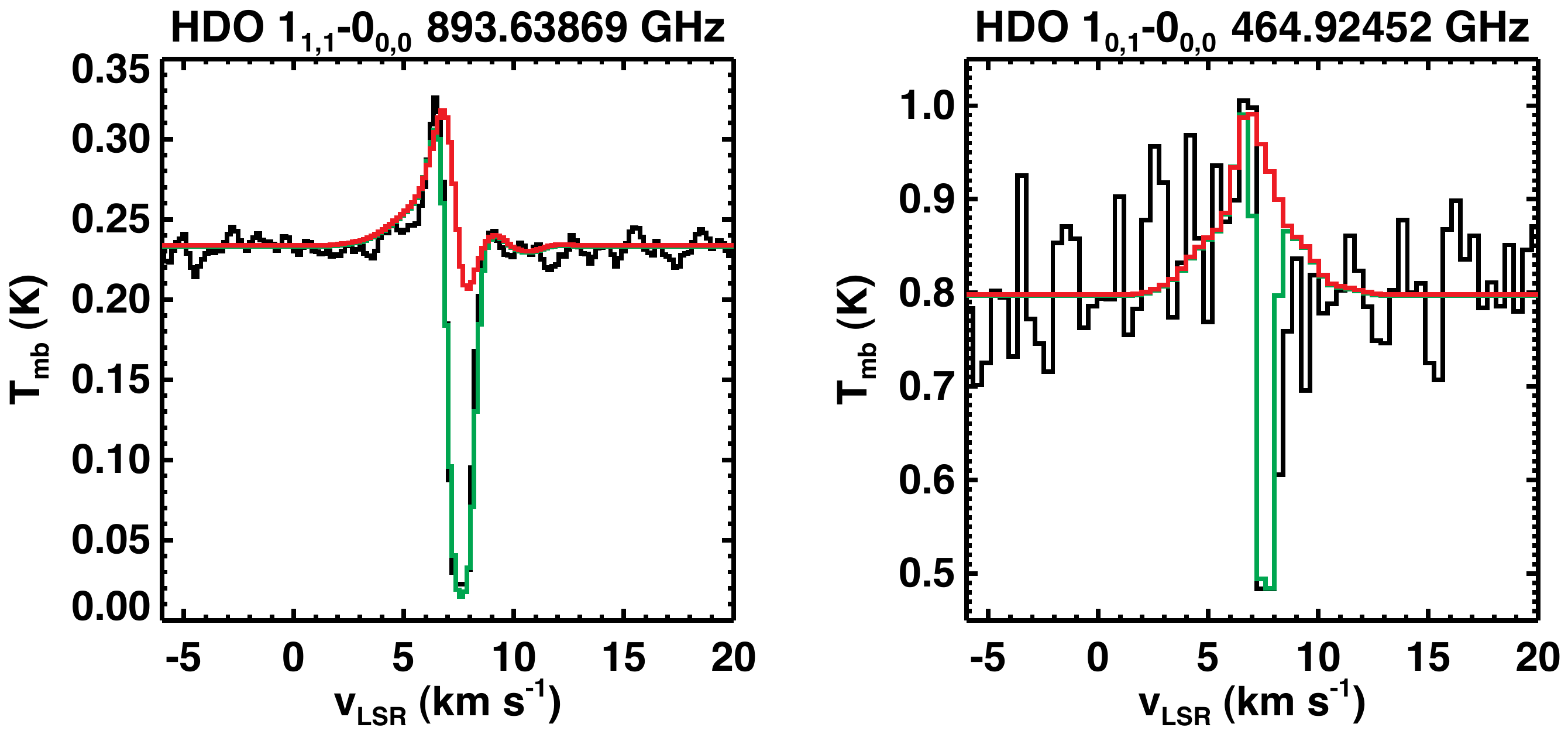}
\caption{Comparison of the modeling of the HDO 1$_{1,1}$--0$_{0,0}$ and 1$_{0,1}$--0$_{0,0}$ fundamental transitions observed towards IRAS~4A with an added absorbing layer (in green) and without this layer (in red). The HDO column density in the absorbing layer is about 1.4 $\times$ 10$^{13}$ cm$^{-3}$. The broad outflow component seen in the observations was subtracted.
The continuum refers to SSB data.}
 \label{comp_abs}
\end{center}
\end{figure}

\begin{figure}[!t]
\begin{center}
\includegraphics[scale=0.52]{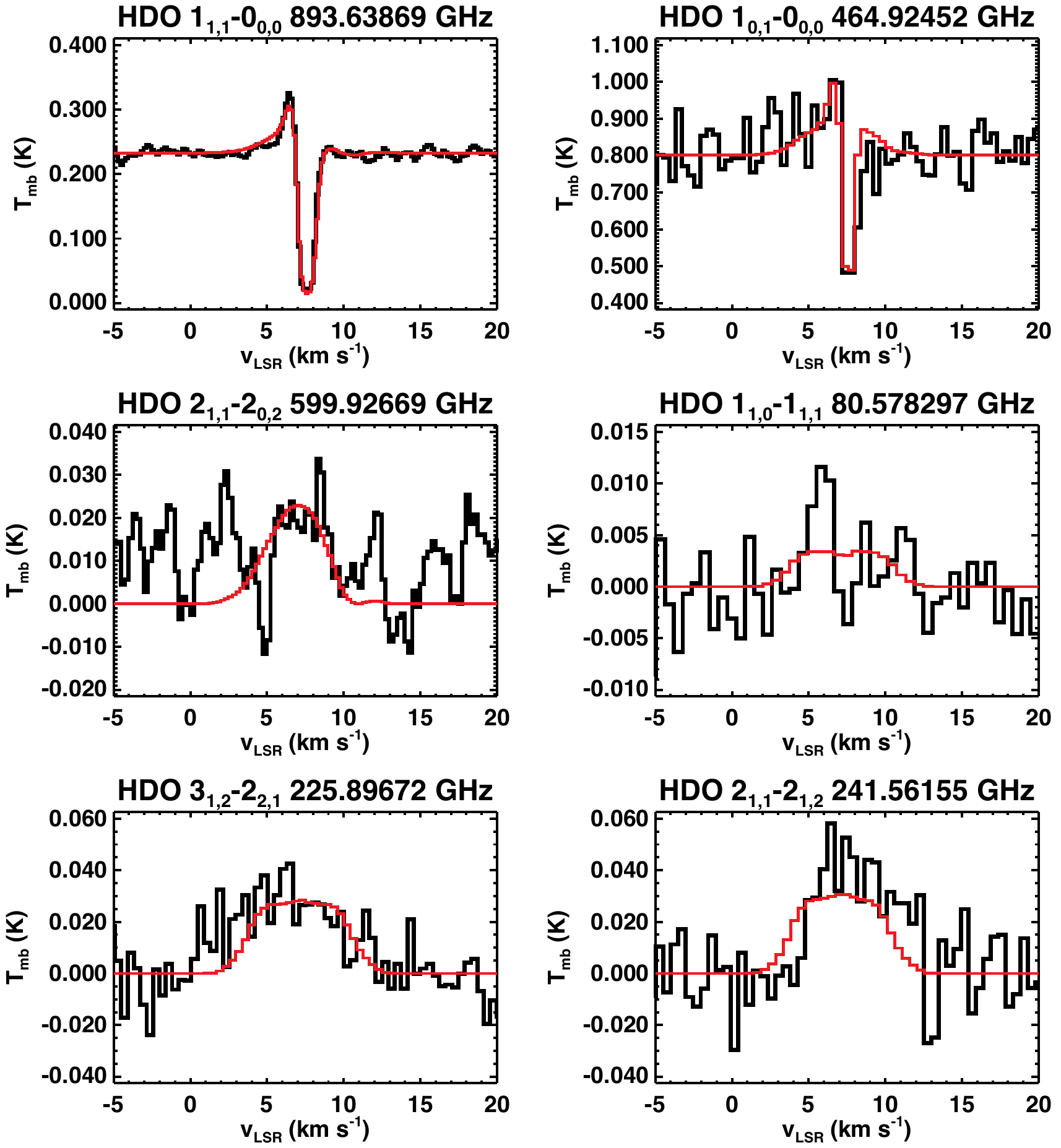}
\caption{In \textit{black}: HDO lines observed towards IRAS~4A with HIFI, IRAM, and JCMT. The broad outflow component seen in the observations was subtracted.
The continuum seen in the observations of the 894 GHz and 465 GHz lines refers to SSB data.
In \textit{red}: best-fit model obtained when adding an absorbing layer with an HDO column density of $\sim$1.4 $\times$ 10$^{13}$ cm$^{-2}$ to the structure.
The best-fit inner abundance is 7.5 $\times$ 10$^{-9}$ and the best-fit outer abundance is 1.2 $\times$ 10$^{-11}$. }
\label{ratran_bestfit_iras4a}
\end{center}
%\end{figure}
%
%\begin{figure}[!ht]
\begin{center}
\includegraphics[scale=0.34]{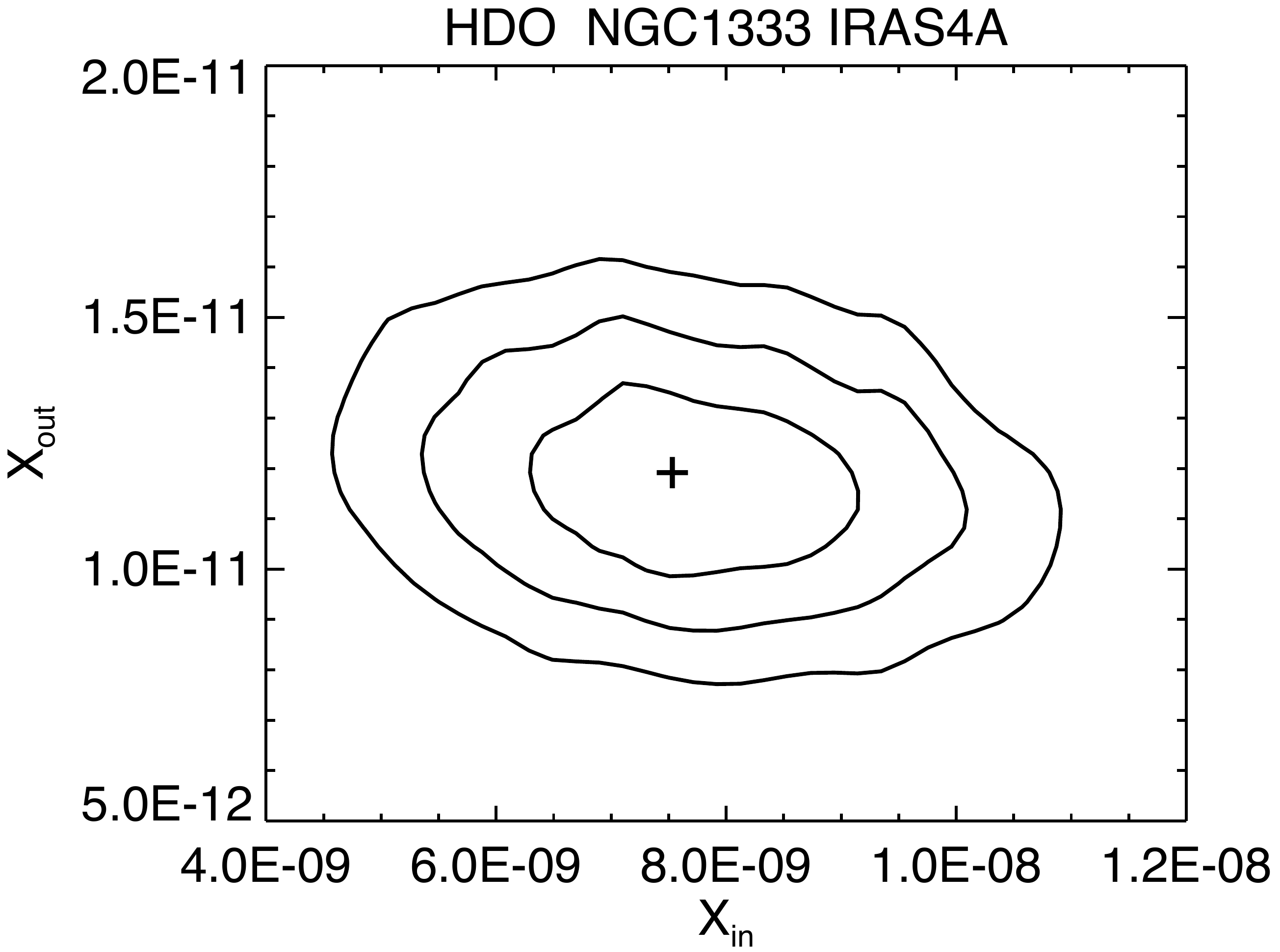}
\caption{$\chi^2$ contours at 1$\sigma$, 2$\sigma$, and 3$\sigma$ obtained for IRAS~4A when adding an absorbing layer with an HDO column density of 1.4 $\times$ 10$^{13}$ cm$^{-2}$ to the structure. The best-fit model is represented by the symbol ``+''. The mass is equal to 0.5 $M_\odot$.}
\label{chi2_iras4a}
\end{center}
\end{figure}

The spherical non-LTE (local thermal equilibrium) radiative transfer code RATRAN \citep{Ratran}, which includes continuum emission and absorption by dust, was used to determine the HDO abundances in the protostellar envelope. 
First, it was necessary to subtract the broad outflow component fitted by a Gaussian profile (see Table \ref{decomp}) from the HDO fundamental transitions at 465 and 894 GHz. 
Then a similar method to the study on the protostar IRAS 16293-2422 \citep{Coutens2012} was carried out on the line profiles obtained after the removal of the broad component. We used the HDO collisional rates calculated with ortho and para--H$_2$ \citep{Faure2011,Wiesenfeld2011}, assuming an ortho/para ratio of H$_2$ at local thermodynamic equilibrium with the gas temperature.
The density and temperature radial profiles of the source IRAS~4A were determined by \citet{Kristensen2012}. The envelope mass is estimated at 5.2 M$_\odot$ \citep{Kristensen2012}, and the radius of the protostellar envelope extends from 33.5 AU to 33500 AU. 
The density power-law index is 1.8. At $r$ = 1000 AU, the density is equal to 6.7 $\times$ 10$^6$ cm$^{-3}$, and the temperature is 21 K.
However, at small scale ($\lesssim$ 500 AU, \citealt{Jorgensen2004b}), the structure is rather uncertain because of the limited spatial resolution of the continuum maps used to determine the profiles. 
The velocity profile is assumed as a free-fall profile (v = $\sqrt{2GM/r}$). The central mass $M$ was estimated at $\sim$ 0.5 $M_\odot$ by \citet{Maret2002} and \citet{Jorgensen2009} and at $\sim$ 0.7 $M_\odot$  by \citet{DiFrancesco2001} and \citet{Mottram2013}. Several grids of models with different values of $M$ (0.3, 0.5, and 0.7 $M_\odot$) were then run. 
To reproduce the absorption component seen in the fundamental lines, the Doppler b-parameter ($db$), which is related to the turbulence broadening, is estimated at 0.4 \kms. 
This is the same as found independently for H$_2$O by \citet{Mottram2013}. It means that the FWHM produced by the turbulence is equal to $db$/0.6, i.e., about 0.67 \kms. 
The continuum is not correctly fitted with the dust opacity used by \citet[][model OH5 in \citealt{Ossenkopf1994}]{Kristensen2012}. It is 17\% lower than the observed continuum at 465 GHz and 28\% higher at 895 GHz. 
Calibration uncertainties could play a role.
The study of the absorbing components requires, however, the use of the correct continuum value.
A power-law emissivity model was then estimated locally to fit the observed continuum,
\begin{equation} 
\kappa = 8.5 \left(\frac{\nu}{10^{12}}\right)^{1.05},
\end{equation}
with $\kappa$  the absorption coefficient in cm$^2$\,g$_{\rm dust}^{-1}$ and $\nu$ the frequency in Hz.
We assumed an abundance profile with a jump at 100 K ($\theta$~$\sim$~0.73$\arcsec$, $r$~$\sim$ 85 AU), for the release by thermal desorption of the water molecules trapped in the icy grain mantles into the gas phase \citep{Fraser2001}.
This type of jump abundance profile was assumed in a number of studies of water and deuterated water in low-mass protostars \citep{Ceccarelli2000,Parise2005,Liu2011,Coutens2012}. A recent study shows, however, that in the outer envelope of low-mass protostars the H$_2$$^{16}$O line profiles are better reproduced with an abundance increasing gradually with radius \citep{Mottram2013}. To compare the HDO abundances in IRAS~4A with previous results in other low-mass protostars and to keep a reasonable number of free parameters in the modeling, we used a jump abundance profile with a constant outer abundance for this study.

Grids of models with different inner ($X_{\rm in}$) and outer ($X_{\rm out}$) abundances\footnote{The HDO abundances quoted in the paper correspond to the N(HDO)/N(H$_2$) ratio.}, and central masses ($M$) were run. 
Similarly to IRAS 16293-2422 \citep{Coutens2012}, an absorbing layer had to be added to the structure to reproduce the absorption lines observed at 894 and 465 GHz, without overpredicting the line emissions. Fig. \ref{comp_abs} shows the difference of the modeling with and without the absorbing layer for an HDO column density of 1.4 $\times$ 10$^{13}$ cm$^{-3}$. The analysis and the discussion regarding this layer are detailed in Sect. \ref{abs_layer}.
A $\chi^2$ minimization was then used to determine the best-fit parameters, adding this absorbing layer to the structure.
The $\chi^2$ is computed on the line profiles with the formalism,
\begin{equation}
 \chi^2 = \sum_{i=1}^{N} \sum_{j=1}^{ n_{\rm chan}} \frac{(T_{{\rm obs},ij}-T_{{\rm mod},ij})^2}{{rms}_{i}^2},
\end{equation}
with $N$ the number of observed lines $i$, $n_{\rm chan}$ the number of channels $j$ for each line, $T_{\rm obs,ij}$ and $T_{\rm mod,ij}$ the intensity observed and predicted by the model in the channel $j$ of the line $i$, and $rms_{ \rm i}$ the rms of the line $i$. 
Taking into account this foreground absorbing layer in the modeling, the best-fit is obtained for an inner abundance \xin =  7.5 $\times$ 10$^{-9}$, an outer abundance \xout =  1.2 $\times$ 10$^{-11}$, and a central mass $M$ = 0.5 $M_\odot$. 
The line profiles predicted by this model are shown in Fig. \ref{ratran_bestfit_iras4a}, and the $\chi^2$ contours at 1, 2, and 3$\sigma$ are plotted in Fig. \ref{chi2_iras4a}. At 3$\sigma$, the inner abundance is between 4.5 $\times$ 10$^{-9}$ and 1.1 $\times$ 10$^{-8}$, whereas the outer abundance is between 8 $\times$ 10$^{-12}$ and 1.6 $\times$ 10$^{-11}$. Although the models with $M$ = 0.3 $M_\odot$ and $M$ = 0.7 $M_\odot$ show higher $\chi^2$ values, some of them are, however, included in the 3$\sigma$ uncertainties. For these masses, the best-fit abundances remain quite similar: \xin = 7.0 $\times$ 10$^{-9}$, \xout =  1.2 $\times$ 10$^{-11}$ for $M$ = 0.3 M$_\odot$ and \xin =  8.5 $\times$ 10$^{-9}$, \xout = 1.0 $\times$ 10$^{-11}$ for $M$ = 0.7 M$_\odot$.

\subsection{The protostellar envelope of NGC~1333 IRAS~4B}
\label{sect_iras4b}

\begin{figure}[!t]
\begin{center}
\includegraphics[scale=0.52]{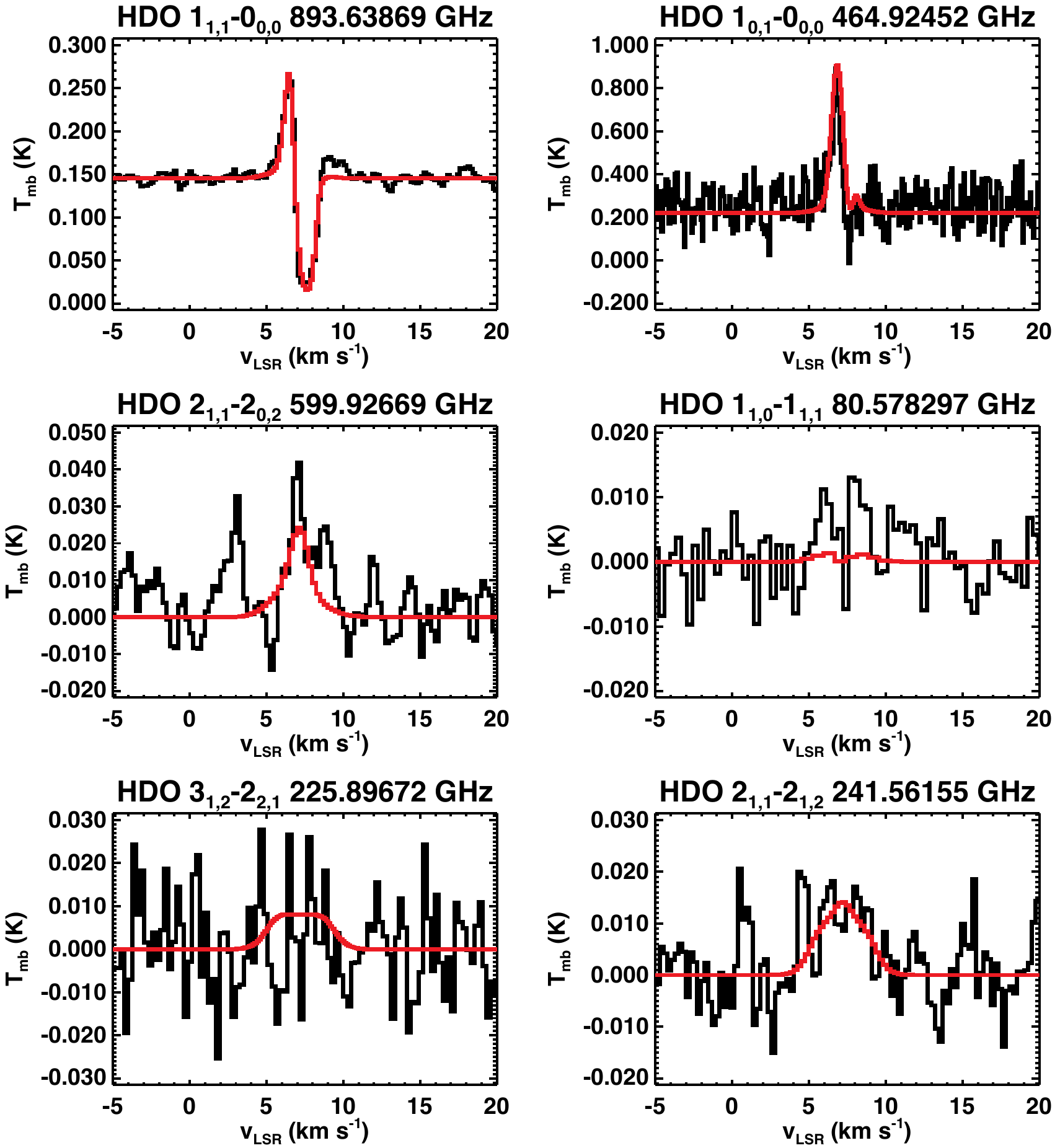}
\caption{In \textit{black}: HDO lines observed towards IRAS~4B with HIFI, IRAM, and APEX. The broad outflow component seen in the observations was subtracted.
The continuum seen in the observations of the 894 GHz and 465 GHz lines refers to SSB data.
In \textit{red}: best-fit model obtained when adding the absorbing layer used to model the IRAS~4A lines (N(HDO) $\sim$1.4 $\times$ 10$^{13}$ cm$^{-2}$). 
The best-fit inner abundance is 1 $\times$ 10$^{-8}$ and the best-fit outer abundance is 1.2 $\times$ 10$^{-10}$. }
\label{ratran_bestfit_iras4b}
\end{center}
%\end{figure}
%
%\begin{figure}[!t]
\begin{center}
\includegraphics[scale=0.32]{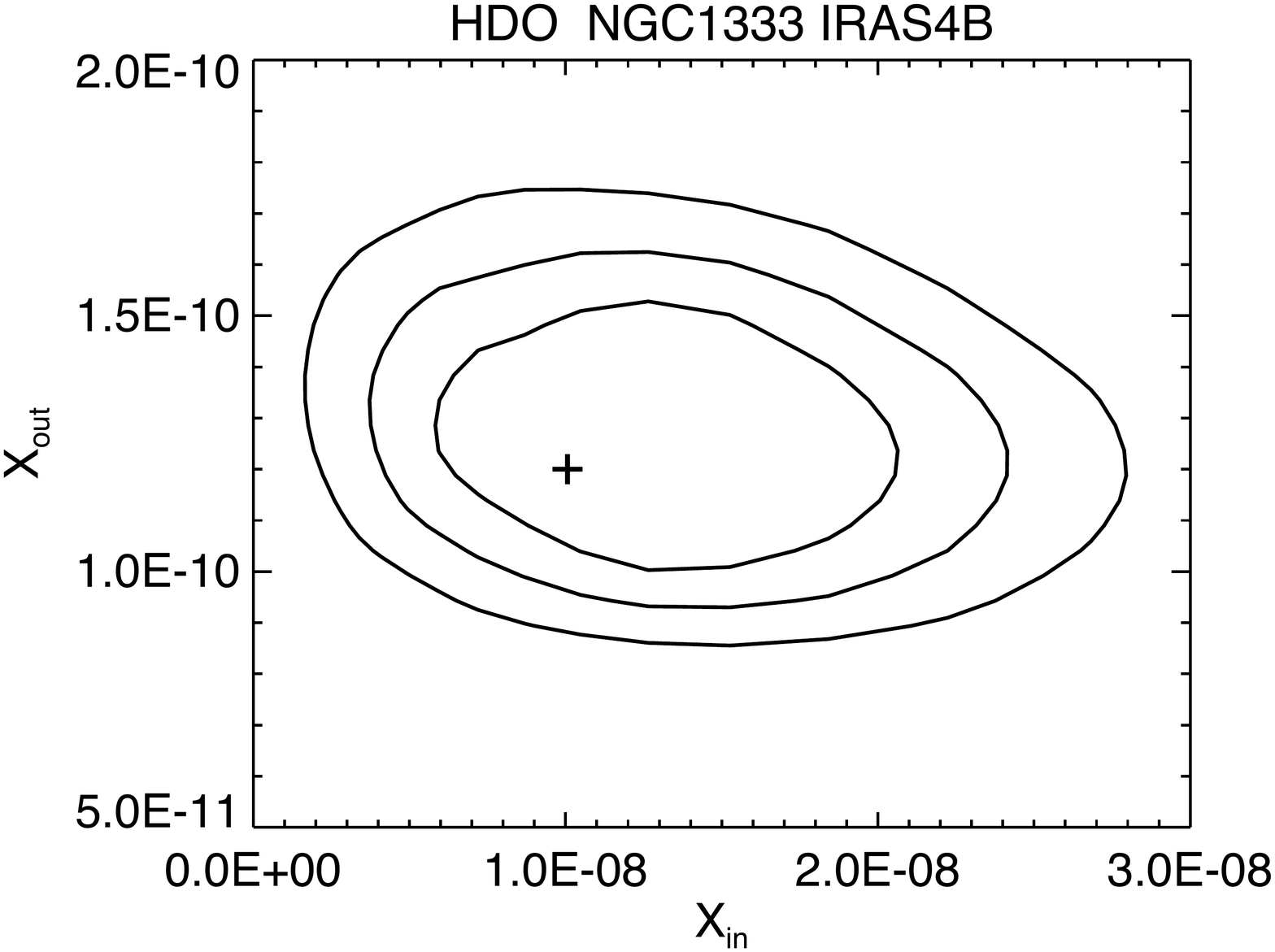}
\caption{$\chi^2$ contours at 1$\sigma$, 2$\sigma$, and 3$\sigma$ obtained for IRAS~4B when adding the absorbing layer used to model the IRAS~4A lines. The best-fit model is represented by the symbol ``+''. The mass is equal to 0.1 $M_\odot$.}
\label{chi2_iras4b}
\end{center}
\end{figure}

A similar modeling was carried out to determine the HDO abundance distribution in IRAS~4B. The outflow components were subtracted here again. We used the source structure determined by \citet{Kristensen2012}, for which the radius of the protostellar envelope ranges between 15 AU and 12000 AU and the envelope mass is about 3.0 M$_\odot$. The density power-law index is 1.4. At $r$ = 1000 AU, the density is 5.7 $\times$ 10$^6$ cm$^{-3}$, and the temperature is 17 K. As in the case of IRAS~4A, the structure is quite uncertain at small scale. According to this structure, the size of the abundance jump (T $>$ 100 K) is $\theta$~$\sim$~0.4$\arcsec$, i.e., $r$~$\sim$ 47 AU.
When the absorbing layer defined for IRAS~4A ($N$(HDO) = 1.4 $\times$ 10$^{13}$ cm$^{-2}$) is added to the structure of its nearby companion IRAS~4B, the absorbing components are well reproduced (see Fig. \ref{ratran_bestfit_iras4b}). The velocity profile is assumed to be a free-fall profile. The central mass $M$ derived in previous studies varies by more than a factor of 2. \citet{DiFrancesco2001} and \citet{Jorgensen2009} predicted a mass about 0.2 $M_\odot$, whereas \citet{Maret2002} determined a mass about 0.5 $M_\odot$. 

Several values were then considered for the central mass $M$ chosen from 0.1 $M_\odot$ to 0.5 $M_\odot$ in steps of 0.1 $M_\odot$. The Doppler b-parameter is estimated at 0.5 \kms, to reproduce the deep absorbing component at 894 GHz.
We used the emissivity model determined by \citet{Ossenkopf1994}, for thin ice-coated grains with a growth of 10$^6$ years (OH5, similar to \citealt{Kristensen2012}), which gives a dust continuum value in agreement with the \textit{Herschel}/HIFI observations. The continuum level at 465 GHz is then predicted at about 0.2 K for the APEX beam.

 \begin{figure*}[!t]
\begin{center}
\includegraphics[scale=0.45]{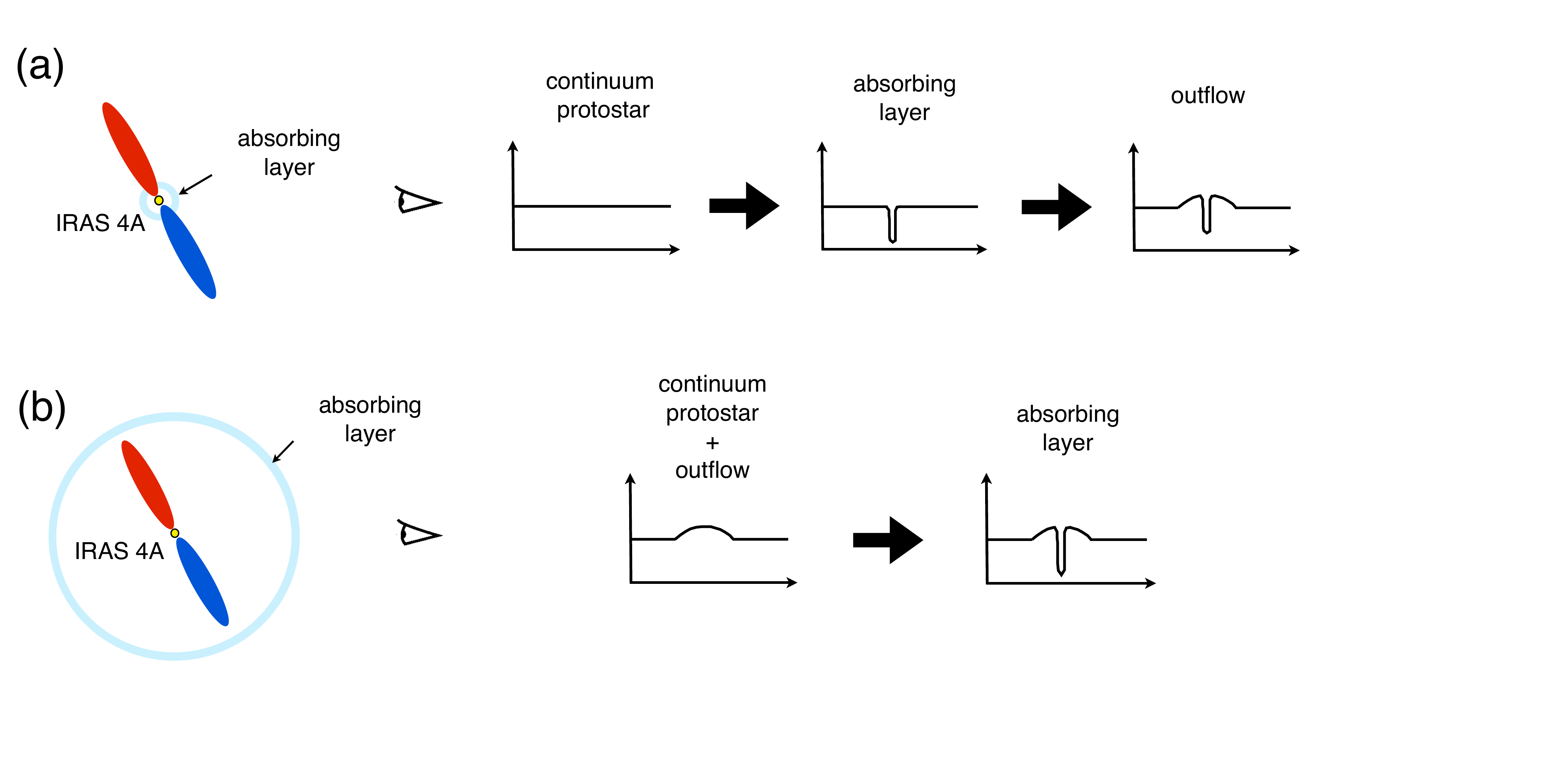}
\caption{Different scenarios for the position of the absorbing layer with respect to the protostar and the outflows and influence on the absorbing line depth. In case (a), the absorbing layer is less extended than the outflows. The continuum that is absorbed is only due to the protostar. The broad component is superposed. In case (b), the absorbing layer is more extended than the outflows. The continuum that is absorbed is due both to the protostar and the outflows.}
 \label{sketch_abs}
\end{center}
\end{figure*} 

Running a grid with different values for $X_{\rm in}$, $X_{\rm out}$, and $M$, the best-fit model is obtained for an inner abundance of about 1 $\times$ 10$^{-8}$, an outer abundance of about 1.2 $\times$ 10$^{-10}$, and a central mass of about 0.1 $M_\odot$.
The best-fit line profiles are overplotted on the observations in Fig. \ref{ratran_bestfit_iras4b}. The $\chi^2$ contours are shown in Fig. \ref{chi2_iras4b}. At 3$\sigma$, the inner abundance varies between 1.5 $\times$ 10$^{-9}$ and 2.8 $\times$ 10$^{-8}$, whereas the outer abundance is included between  8.5 $\times$ 10$^{-11}$ and 1.75 $\times$ 10$^{-10}$.
Some models with $M$ = 0.2 $M_\odot$ are also included in the 3$\sigma$ uncertainties, but not for masses higher than or equal to 0.3 $M_\odot$.

\subsection{The absorbing layer}
\label{abs_layer}

Depending on the position of the absorbing layer with respect to the outflows, the depth of the absorption components may vary after the subtraction of the broad component.
Indeed, if the absorbing layer is situated between the outflows and the protostellar envelope (see case (a) in Fig. \ref{sketch_abs}), the depth of the absorption line has to be unchanged after the subtraction of the large component, because the continuum produced by the outflows (i.e., the broad component) is not absorbed by the added layer. On the contrary, if the absorbing layer is located in front of the outflows (see case (b) in Fig. \ref{sketch_abs}), the continuum obtained after subtraction is lower than the real absorbed continuum, since the line wing effectively contributes to the continuum that is being absorbed.
In this case, to accurately estimate the HDO column density in the absorbing layer, the 
depths of the absorbing lines must be corrected, reducing them by a value close to the intensities of the broad components. The uncertainty on this value is $\lesssim$ 10\%, which is comparable to the calibration uncertainties of the HIFI observations and is probably even lower for the JCMT observations.
The presence of an absorption component in the data at 894 GHz towards the outflow of IRAS~4A (see Fig. \ref{hdo_outflow}) suggests that the foreground absorbing layer is more extended than a part of the outflows (case (b) in Fig. \ref{sketch_abs}). 
When subtracting the broad components on the IRAS~4A observations, the depth of the absorbing lines was consequently reduced to take into account the absorption by the continuum produced by the outflows.
We assumed a similar behavior for IRAS~4B and treated the absorptions in the same way. 

To reproduce the depth of the absorption lines, the column density of HDO is then about 1.4 $\times$ 10$^{13}$ cm$^{-2}$. The temperature is assumed to be between 10 and 30 K and the H$_2$ density is found to be lower than 10$^5$ cm$^{-3}$ to reproduce the absorptions. 
This upper limit on the H$_2$ density is consistent with the study by \citet{Coutens2013} using the D$_2$O lines observed towards IRAS~16293-2422.
The HDO column density derived here differs by less than a factor of 2 compared with the estimate in the absorbing layer of IRAS 16293-2422 \citep[2.3 $\times$ 10$^{13}$ cm$^{-2}$;][]{Coutens2012}. 
As the absorbing layer shows the same characteristics for IRAS~4A and IRAS~4B, this could mean that this layer is sufficiently extended ($\gtrsim$ 7000 AU) to encompass the two sources.

Several molecules (CO, HCN, HCO$^+$, H$_2$CO) show, over a large extent ($\gtrsim$ 200$\arcsec$) around IRAS~4, asymmetric line-profiles with dips at a velocity about 8 $\kms$ \citep[$\sim$ 7.5 $\kms$ at the IRAS~4A position;][]{Choi2004}. 
In our data, the velocity of the absorbing components appears at 7.6 \kms. These molecules and water could consequently arise from the same absorbing layer. 
The origin of this absorbing layer was largely debated \citep[e.g.,][and references therein]{Langer1996,DiFrancesco2001,Choi2001}. In particular, \citet{Choi2004} interpreted these absorptions as produced by a cold foreground layer unrelated to the IRAS~4A source.
This interpretation was, however, questioned by \citet{Belloche2006}, as they showed that an infall motion would be sufficient to explain the observed absorptions of CS and N$_2$H$^+$ transitions if a source velocity of 7.2 \kms~is assumed instead of the value of 6.7 \kms~used by \citet{Choi2004}. \citet{Walsh2006} drew similar conclusions, favoring the interpretation that the HCO$^+$ line profiles are indicative of large-scale inward motions associated with the IRAS4A source.
In our study, nothing allows us to conclude that the water-rich absorbing layer is unrelated to the protostellar sources. This cold layer could consequently be a part of the complex surrounding IRAS~4.

While it is not surprising to detect molecules such as CO, HCO$^+$, etc., in cold environments, the presence of water is more puzzling. Indeed, at low temperatures, water is mainly trapped on the grain mantles.
Hypotheses were suggested by \citet{Coutens2012} to explain the presence of a water-rich absorbing layer surrounding the protostar IRAS~16293-2422.
The most probable explanation is that this layer results from an equilibrium between the photodesorption of water molecules trapped in the icy grain mantles and their photodissociation by an external UV radiation field, as predicted by \citet{Hollenbach2009} in molecular clouds at a visual extinction $A_{\rm V}$ between 1 and 4. Assuming the relation $N_{\rm H}/$$A_{\rm V}$ $= 2 \times 10^{21}$ cm$^{-2}$ mag$^{-1}$ \citep{Vuong2003}, the deuterated water abundance with respect to H$_2$ would then be $\sim$3.7 $\times$ 10$^{-9}$ -- 1.5 $\times$ 10$^{-8}$ in the assumed photodesorption layer of IRAS~4, which is comparable to the layer surrounding IRAS~16293-2422 ($\sim$ 6 $\times$ 10$^{-9}$ -- 2.4 $\times$ 10$^{-8}$).
Other mechanisms able to explain the presence of gaseous water in a cold layer are discussed in Appendix \ref{sect_sio}.

 We have to point out that photodesorption by the cosmic-ray induced UV-field certainly plays a role in the outer envelope, as shown by \citet{Mottram2013} in low-mass protostars and \citet{Caselli2012} in prestellar cores.
In this case, the abundance increases gradually with the outer radius. 
As the outer abundance is assumed to be constant here, a part of the absorption could also be produced in the outermost part of the envelope (which encompasses the outflow position and shows low temperatures as well as low densities). The column density of water derived in the added absorbing layer should then be considered to be an upper limit.
An analysis of the HDO lines with a more physical abundance profile in the outer layer would be required to know the proportion of cold water in the outer envelope and at larger scale in the molecular cloud.

\subsection{Outflow of NGC~1333 IRAS~4A}
\label{sect_outflow}

\begin{figure}[!t]
\begin{center}
\includegraphics[scale=0.40]{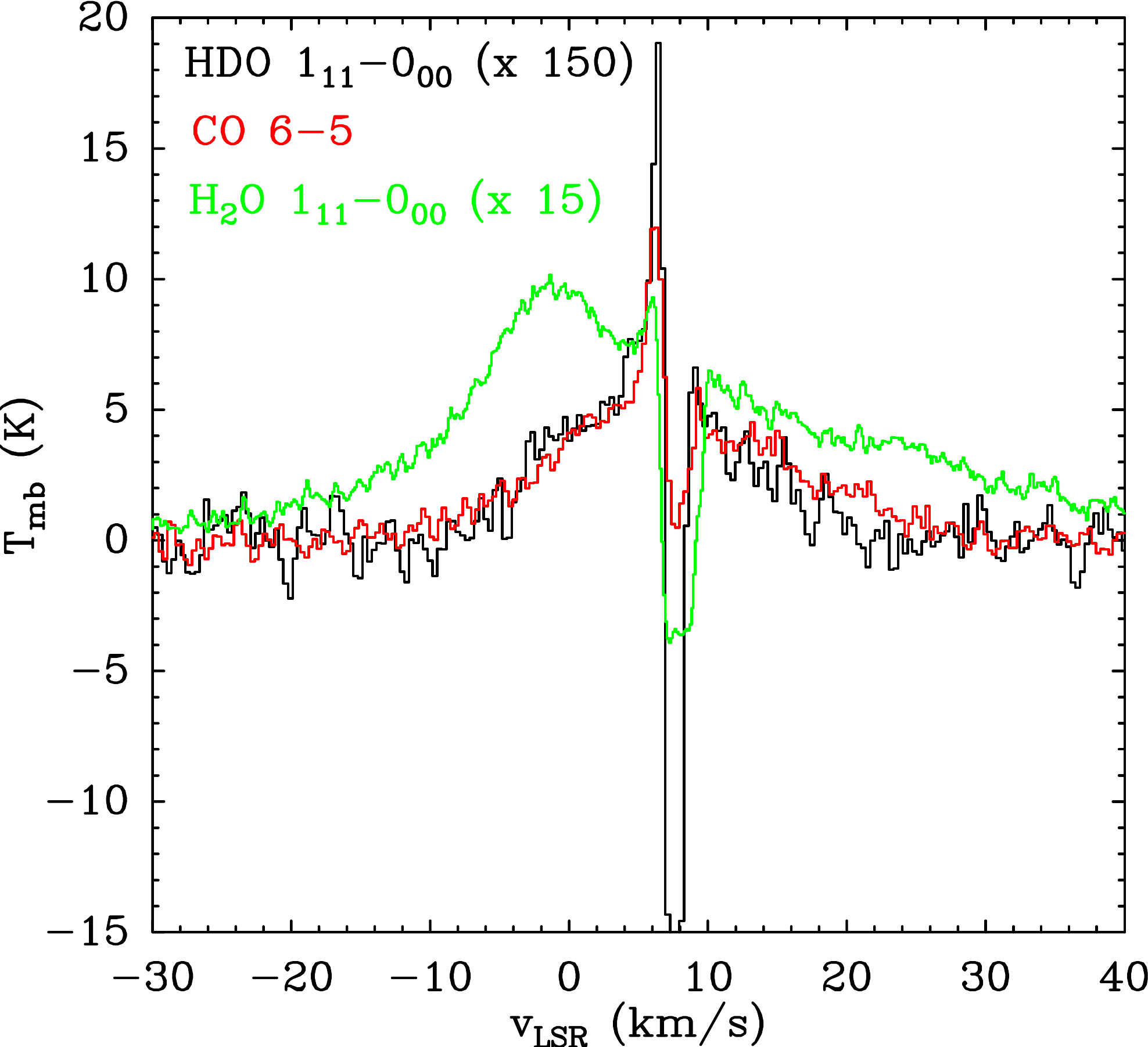}
\caption{Comparison of the line profiles of the CO 6-5 transition (in red), the HDO 1$_{1,1}$--0$_{0,0}$ transition (in black, intensities multiplied by 150), and the H$_2$O 1$_{1,1}$--0$_{0,0}$ transition (in green, intensities multiplied by 15) towards IRAS~4A.}
\label{co_comp}
\end{center}
\end{figure}

A broad component tracing the outflows is seen in the two fundamental 1$_{1,1}$--0$_{0,0}$ and 1$_{0,1}$--0$_{0,0}$ transitions observed with HIFI and JCMT towards IRAS~4A. This HDO outflow component is very similar to the broad component seen on the CO 6--5 line profile observed with APEX by \citet{Yildiz2012} (see Fig. \ref{co_comp}), suggesting a common origin for the entrainment of the HDO and CO molecules. 
The RADEX non-LTE radiative transfer code \citep{vanderTak2007} was employed to estimate the HDO column density present in the outflows of IRAS~4A. We used a H$_2$ density of 3 $\times$ 10$^5$ cm$^{-3}$ and a temperature between 100 and 150 K, as determined by \citet{Yildiz2012} for CO in the outflow positions R1 and B1. The HDO column density is estimated at $\sim$ (2--4) $\times$ 10$^{13}$ cm$^{-2}$. These estimates remain valid, even if the temperature is lower than 100 K. 
Using the H$_2$ column density of (2.1--2.8) $\times$ 10$^{22}$ cm$^{-2}$ estimated by \citet{Yildiz2012} in the outflows, the HDO abundance is therefore about 7 $\times$ 10$^{-10}$ -- 1.9 $\times$ 10$^{-9}$.

The HDO/H$_2$O ratio in the outflows can be also estimated with RADEX using the line ratios of the HDO 1$_{1,1}$--0$_{0,0}$ line at 894 GHz with the para--H$_2$O 1$_{1,1}$--0$_{0,0}$ line at 1113 GHz \citep{Kristensen2010} . At these frequencies, the beam sizes are quite similar (19$\arcsec$ vs 24$\arcsec$).
An ortho-to-para ratio of 3 is assumed to consider the H$_2$O abundance. 
The profile of the para--H$_2$O line appears quite different from that of the HDO line (see Fig. \ref{co_comp}). The H$_2$O outflow profile shows a peak intensity at $-0.6$ $\kms$ that we do not see for the HDO lines. \citet{Kristensen2010} fitted it with a sum of two Gaussians, one centered at $-0.6$ $\kms$ and a broader one centered at 8.7 $\kms$. However, it is not possible to reproduce the HDO profile with the same velocity and FWHM. 
To estimate the HDO/H$_2$O abundance ratio in the outflows, we consequently compared the predicted line ratios with the line ratios at the velocities where the HDO broad component is detected.
A  range of densities (10$^5$--10$^7$ cm$^{-3}$) and temperatures (100--1000 K) was considered.
The derived HDO/H$_2$O ratio ranges between 1 $\times$ 10$^{-3}$ and 9 $\times$ 10$^{-2}$ in the red part of the outflow and between 7 $\times$ 10$^{-4}$ and 6 $\times$ 10$^{-2}$ in the blue part of the outflow.
These estimates are consistent with the HDO/H$_2$O ratio derived by \citet{Taquet2013} in the hot corino of IRAS~4A (5 $\times$ 10$^{-3}$ -- 3 $\times$ 10$^{-2}$).  
If the HDO/H$_2$O ratios are actually similar in the outflow and in the hot corino, it could mean that water is contained in grain mantles and then released into the gas phase by sputtering in outflows and thermal desorption in the warm inner envelope.
The isotopologue HDO would then be omnipresent and produced early in the evolution of the core.
However, the wide range of values determined here does not allow us to clearly assert this possibility.
The HDO/H$_2$O ratio is not determined at high velocities ($\Delta \varv$ $>$ 10 $\kms$) where only H$_2$O is detected. 

\section{Discussion}

\subsection{Comparison of the HDO abundances in low-mass star-forming regions}

\begin{table*}[!ht]
\renewcommand{\arraystretch}{1.2}
\caption{HDO abundances with respect to H$_2$ determined in several low-mass star-forming regions }
\begin{center}
\begin{tabular}{l|c|c|c|c|l}
\hline \hline
Source & Inner  & Outer  & Absorbing layer & Outflow & References \\
& abundance$^{(a)}$ & abundance$^{(a)}$ & N(HDO) (cm$^{-2}$) &  abundance & \\
\hline
NGC 1333 IRAS~4A &  7.5$^{+3.5}_{-3.0}$ $\times$ 10$^{-9}$ & 1.2$^{+0.4}_{-0.4}$ $\times$ 10$^{-11}$ & 1.4 $\times$ 10$^{13}$ & 0.7 -- 1.9 $\times$ 10$^{-9}$ & This paper \\
NGC 1333 IRAS~4B & 1.0$^{+1.8}_{-0.9}$ $\times$ 10$^{-8}$ & 1.2$^{+0.6}_{-0.4}$  $\times$ 10$^{-10}$ &  1.4 $\times$ 10$^{13}$  & -- &  This paper \\
NGC 1333 IRAS~2A & 8$^{+2.0}_{-1.2}$ $\times$ 10$^{-8}$ & 7$^{+11}_{-6.1}$ $\times$ 10$^{-10}$ & -- & -- & \citet{Liu2011} \\
IRAS~16293-2422 & 1.8$^{+0.6}_{-0.4}$ $\times$ 10$^{-7}$ & 8$^{+2.0}_{-3.4}$ $\times$ 10$^{-11}$ & 2.3 $\times$ 10$^{13}$ & -- & \citet{Coutens2012,Coutens2013} \\
L1448-mm & 4 $\times$ 10$^{-7}$ & -- & -- & -- & \citet{Codella2010} \\
\hline
\end{tabular}
\end{center}
\label{conclu_abondances}
\begin{small}
$^{(a)}$ The uncertainties correspond to the 3$\sigma$ confidence interval.
\end{small}
\end{table*}%

\begin{figure*}[!ht]
\begin{center}
\includegraphics[scale=0.7]{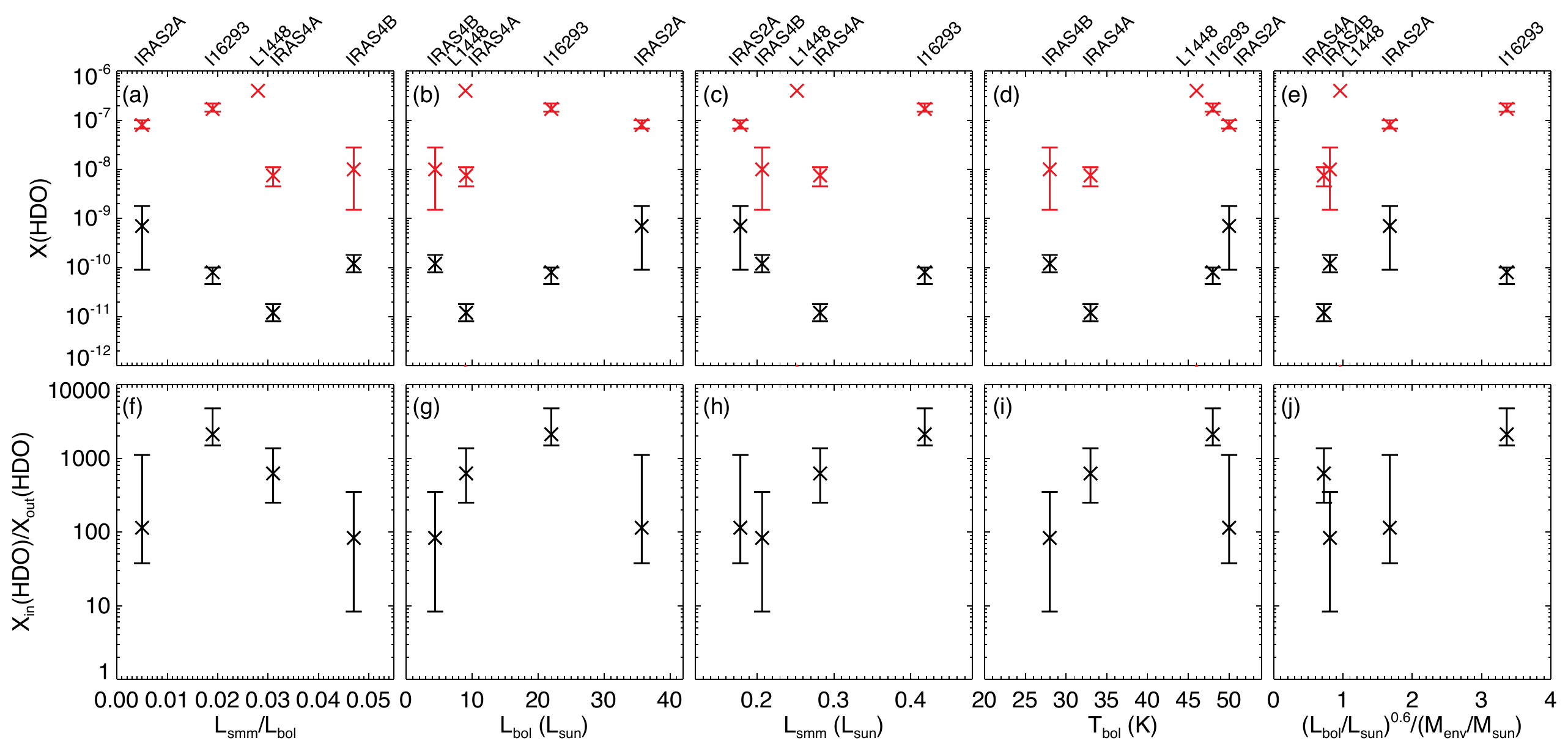}
\caption{Upper panels: Comparison of the inner (red) and outer (black) HDO abundances estimated in the low-mass protostars IRAS~16293-2422, NGC~1333 IRAS~4A, NGC~1333 IRAS~2A, and L1448-mm (see also Table \ref{conclu_abondances}) as a function of the ratio between the submillimeter and bolometric luminosities $L_{\rm smm}$/$L_{\rm bol}$ (a); the bolometric luminosity $L_{\rm bol}$ (b); the submillimeter luminosity $L_{\rm smm}$ (c); the bolometric temperature (d); and the ratio $L_{\rm bol}$$^{0.6}$/$M_{\rm env}$ (e). 
Lower panels: Comparison of the ratio between the inner and outer HDO abundances $X_{\rm in}$/$X_{\rm out}$ as a function of the $L_{\rm smm}$/$L_{\rm bol}$ ratio (f); $L_{\rm bol}$ (g); $L_{\rm smm}$ (h); $T_{\rm bol}$ (i); and $L_{\rm bol}$$^{0.6}$/$M_{\rm env}$ (j). 
The different values of the parameters ($L_{\rm smm}$/$L_{\rm bol}$, $L_{\rm bol}$, $L_{\rm smm}$, $T_{\rm bol}$, and $M_{\rm env}$) come from \citet{Kristensen2012}, \citet{Karska2013}, \citet{Evans2009}, \citet{Froebrich2005}, and \citet{Crimier2010}. The HDO abundance is only estimated in the inner part of the protostellar envelope of L1448-mm using interferometric data \citep{Codella2010}. No uncertainty is provided for this value.}
\label{comp_hdo_xin_xou}
\end{center}
\end{figure*}

Table \ref{conclu_abondances} summarizes the HDO abundances measured in low-mass protostars. 
For all sources except L1448-mm, the derived abundances are based on RATRAN modeling of multi-line single-dish observations. For IRAS~16293-2422, IRAS~4A, and IRAS~4B, the HDO fundamental 1$_{1,1}$--0$_{0,0}$ transition was observed at high sensitivity with \textit{Herschel}/HIFI, allowing us to strongly constrain the outer abundance of deuterated water.
The estimate of the HDO inner abundance in L1448-mm by \citet{Codella2010} is based on interferometric observations of the HDO $1_{1,0}$--1$_{1,1}$ line at 80.6 GHz. Using the spherical source structure determined by \citet{Jorgensen2002}, the abundance is then estimated at 4 $\times$ 10$^{-7}$. The source IRAS~4A shows the lowest HDO abundances among the low-mass protostars (both inner and outer), whereas IRAS~16293-2422 shows the highest inner abundance and IRAS~2A the highest outer abundance.
The HDO outer abundances do not cover a wide range of values. They only range between $\sim$1 $\times$ 10$^{-11}$ and a few 10$^{-10}$. In addition, the HDO outer abundance of IRAS~2A ($\sim$7 $\times$ 10$^{-10}$) is not as strongly constrained as in the other sources, because the modeling did not include the HDO line observed with HIFI at 894 GHz. At 3$\sigma$, the HDO abundance could be as low as 9 $\times$ 10$^{-11}$ in the colder envelope.

We searched for a correlation between the HDO abundances and different source parameters such as the submillimeter luminosity $L_{\rm smm}$ (i.e., measured longward of 350 $\mu$m), the bolometric luminosity $L_{\rm bol}$, and the bolometric temperature $T_{\rm bol}$. 
Fig. \ref{comp_hdo_xin_xou} shows the inner and outer abundances as a function of these parameters as well as the $L_{\rm smm}$/$L_{\rm bol}$ and $L_{\rm bol}$$^{0.6}$/$M_{\rm env}$ ratios (where $M_{\rm env}$ is the envelope mass).
The $L_{\rm smm}$/$L_{\rm bol}$ ratio is an indicator of the evolution stage of star formation \citep{Andre1993}.  The higher it is, the less evolved the protostar is.
The function $L_{\rm bol}$$^{0.6}$/$M_{\rm env}$ is also considered to be correlated with the protostellar evolution \citep{Bontemps1996}.
The bolometric temperature is a third parameter used to evaluate the evolutionary stage of the protostar \citep{Myers1993}, a higher bolometric temperature meaning that the source is more evolved. %, the older the source is.
We do not find correlations of the abundances with the evolution of the protostars, $L_{\rm smm}$ or $L_{\rm bol}$;
however we have  to be careful with the estimate of the inner abundances. 
They are derived using the density profile estimated at large scale. The structure is consequently not constrained at scales comparable to the hot corino size. In addition, it seems that disks could already be present at the Class 0 stage \citep[see for example][]{Pineda2012}.
The ratio between the inner and outer abundances, also plotted in Fig. \ref{comp_hdo_xin_xou}, could be correlated with the submillimeter luminosity which is physically related to the envelope mass.
However, it is not clear why this would happen. This correlation could be simply fortuitous. The low number of studies taken into account does not allow us to assert it at the moment. 

Another point to check is the variation of the HDO abundances with the physical environment. Indeed the HDO abundances could also depend on the initial conditions of the molecular cloud in which they are embedded. In this case, it would be sensible to only compare the results for the sources situated in the NGC~1333 complex. A correlation between the HDO outer abundances and the submillimeter luminosities of the three NGC~1333 sources is then observed. The lower the $L_{\rm smm}$ parameter, the higher the outer abundance, which could simply reflect the fact that water trapped in the icy grain mantles is more efficiently released in the gas phase by non-thermal desorption mechanisms when the density is less important (i.e., when $L_{\rm smm}$ is low). Cosmic rays and UV photons could then penetrate more deeply in the envelope to desorb water molecules. It is also possible that it favors the ion-molecule reactions in the gas phase. This hypothesis is strengthened by the comparison of the outer abundance with the value of the density at 1000 AU (1.5 $\times$ 10$^6$, 5.7 $\times$ 10$^6$, and 6.7 $\times$ 10$^6$ cm$^{-3}$ for IRAS~2A, IRAS~4B, and IRAS~4A, respectively), as $X_{\rm out}$ increases with decreasing H$_2$ densities. Therefore, the difference between the submillimeter luminosities of IRAS~4A and IRAS~4B could explain why the outer abundance is higher in IRAS~4B.

\subsection{Water deuterium fractionation}
\label{sect_ratio}

\begin{table*}[!ht]
\renewcommand{\arraystretch}{1.2}
\caption{Summary of the HDO/H$_2$O ratios determined in Class 0 protostars.}
\begin{center}
\begin{tabular}{c|c|c|c|l|c}
\hline \hline
Source & Study$^{(1)}$ & Inner HDO/H$_2$O ratio & Outer HDO/H$_2$O ratio & Observations$^{(2)}$ & Modeling$^{(3)}$ \\
\hline
IRAS~16293-2422 & a & 3 $\times$ 10$^{-2}$ & $\leq$ 2 $\times$ 10$^{-3}$ & HDO: IRAM (4), JCMT (1) & 1D, non-LTE \\
& & & & H$_2$$^{16}$O: ISO (11) & \\
\cline{2-6}
 & b & 4 $\times$ 10$^{-3}$ -- 5.1 $\times$ 10$^{-2}$ & 3 $\times$ 10$^{-3}$ -- 1.5 $\times$ 10$^{-2}$ & HDO: HIFI (9), IRAM (3), JCMT (1) & 1D, non-LTE \\
 & & & & H$_2$$^{18}$O: HIFI (5) & \\
 \cline{2-6}
 & c & (9.2 $\pm$ 2.6) $\times$ 10$^{-4}$ & -- & HDO: SMA (1) & 0D, LTE\\
  & & & & H$_2$$^{18}$O: ALMA (1), SMA (1) & \\
\hline
NGC1333 IRAS2A & d & $\geq$ 1 $\times$ 10$^{-2}$ &  9 $\times$ 10$^{-3}$ -- 1.8 $\times$ 10$^{-1}$ & HDO: IRAM (3), JCMT (1), APEX (1) & 1D, non-LTE\\
  & & & & H$_2$$^{18}$O: HIFI$^*$ (non-detection) & \\
    & & & & H$_2$$^{16}$O: HIFI$^*$ & \\
 \cline{2-6}
& e & $\sim$ 1 $\times$ 10$^{-3}$ & -- & HDO: IRAM (1), & 0D, LTE \\
  & & & & H$_2$$^{18}$O: PdBi (1), HIFI (1) & \\
 \cline{2-6}
& f & 3 $\times$ 10$^{-3}$ -- 8 $\times$ 10$^{-2}$ & -- & HDO: PdBi (1), IRAM (3) & 0D, non-LTE \\
  & & & & H$_2$$^{18}$O: PdBi (1) & \\
\hline
NGC1333 IRAS4A & f & 5 $\times$ 10$^{-3}$ -- 3 $\times$ 10$^{-2}$ & -- & HDO: PdBi (1) & 0D, non-LTE \\
  & & & & H$_2$$^{18}$O: PdBi (1) & \\
 \cline{2-6}
& This & 4 $\times$ 10$^{-4}$ -- 3.0 $\times$ 10$^{-3}$ & about 10$^{-2}$ $^{(4)}$  & HDO: HIFI (1), IRAM (3), JCMT (1) & 1D, non-LTE \\
& paper & & & H$_2$$^{18}$O: PdBi$^*$ (1) & \\
&  & & & H$_2$$^{16}$O: HIFI$^*$ (4) & \\
 \hline
NGC1333 IRAS4B & g & $\leq$ 6 $\times$ 10$^{-4}$ & -- & HDO: SMA (non-detection) & 0D, non-LTE \\
  & & & & H$_2$$^{18}$O: PdBi (1) & \\
 \cline{2-6}
& This & 1 $\times$ 10$^{-4}$ -- 3.7 $\times$ 10$^{-3}$ & -- & HDO: HIFI (2), IRAM (2), APEX (1) & 1D, non-LTE\\
& paper & & & H$_2$$^{18}$O: PdBi$^*$ (1) & \\
\hline
\end{tabular}
\end{center}
\label{comp_ratios}
\begin{small}
$^{(1)}$ References classified by order of publication for each source: a) \citet{Parise2005,Ceccarelli2000}; b) \citet{Coutens2012,Coutens2013}; c) \citet{Persson2013}; d) \citet{Liu2011}; e) \citet{Visser2013}; f) \citet{Taquet2013}; g) \citet{Jorgensen2010}.  \\
$^{(2)}$ This column shows which tracers are used to determine the HDO/H$_2$O ratios, which telescopes the analysis is based on, and how many detected lines (number into brackets) of each tracer are used. 
The notation IRAM refers to the single-dish 30m-telescope. The star ($^*$) shows that the modeling of the water lines was not carried out in the study quoted in column 2, but come, as explained in the text, from the following papers: \citet{Jorgensen2010,Persson2012,Taquet2013,Mottram2013}. \\
$^{(3)}$ This column describes the type of modeling used to determine the HDO/H$_2$O ratio: 0D or 1D structure and LTE or non-LTE analysis.\\ 
$^{(4)}$ This value has to be confirmed using a HDO abundance profile similar to the H$_2$O profile in \citet{Mottram2013}.
\end{small}
\end{table*}%

\subsubsection{The warm inner HDO/H$_2$O ratios}

The warm HDO/H$_2$O ratio cannot be directly determined by comparing $X_{\rm in}$(HDO) with the abundance of H$_2$$^{18}$O derived by \citet{Persson2012}, as the H$_2$ densities are estimated in different ways. It is, however, possible to estimate it using the HDO column density derived with our modeling in the warm inner regions of the envelope and the column density of water derived with interferometric observations of the para--H$_2$$^{18}$O 3$_{1,3}$--2$_{2,0}$ transition at 203 GHz.

Including the 3$\sigma$ uncertainties, we obtain a column density of HDO of about (1.4 -- 3.5) $\times$ 10$^{16}$ cm$^{-2}$ in the warm inner envelope (T $\geq$ 100 K) of IRAS~4A.
The column density of p--H$_2$$^{18}$O is estimated by \citet{Taquet2013} to be between 5.5 $\times$ 10$^{15}$ and 1.5 $\times$ 10$^{16}$ cm$^{-2}$ for a hot corino size comparable to ours ($\sim$ 0.73$\arcsec$, $\sim$ 85 AU). It is also consistent with the estimate of 7.9 $\times$ 10$^{15}$ cm$^{-2}$ by \citet{Persson2012} for the same object.
Assuming an ortho-to-para ratio of water equal to 3 and an isotopic H$_2$$^{16}$O/H$_2$$^{18}$O ratio of 500, we can consequently estimate the inner HDO/H$_2$O ratio to be about 4 $\times$ 10$^{-4}$ -- 3.0 $\times$ 10$^{-3}$ in the warm envelope of IRAS~4A. This result is, however, lower than the estimate by \citet{Taquet2013} of 5 $\times$ 10$^{-3}$ -- 3 $\times$ 10$^{-2}$.
In IRAS~4B, we find a column density of deuterated water between 1.0 $\times$ 10$^{15}$ and 3.0 $\times$ 10$^{16}$ cm$^{-2}$.
Using the para--H$_2$$^{18}$O column density estimated by \citet{Jorgensen2010b} at 4 $\times$ 10$^{15}$ cm$^{-2}$, the HDO/H$_2$O ratio is then about 1 $\times$ 10$^{-4}$ -- 3.7 $\times$ 10$^{-3}$. A part of this range is consequently in agreement with the upper limit previously determined by \citet[][ $\leq$ 6 $\times$ 10$^{-4}$]{Jorgensen2010}. The range in HDO/H$_2$O determined for IRAS~4B appears larger than for IRAS~4A because of the larger uncertainties obtained with the HDO inner abundances.

Table \ref{comp_ratios} summarizes the HDO/H$_2$O ratios derived in Class 0 sources in this paper and in previous studies.
The results found here seem to favor low values around 10$^{-3}$ for the warm HDO/H$_2$O ratios. We have to keep in mind that in our modeling, the HDO inner abundance is constrained thanks to the HDO excited lines at 226 and 242 GHz. In \citet{Jorgensen2010}, the non-detection of the line at 226 GHz with the SMA was used to determine an upper limit on the water D/H ratio. The use of the same lines explains easily why our results are in agreement with \citet{Jorgensen2010}. \citet{Taquet2013} used interferometric observations of the more excited HDO 4$_{2,2}$--4$_{2,3}$ line at 144 GHz ($E_{\rm up}$ = 319 K) to derive the HDO/H$_2$O ratio using a non-LTE approach. These different results suggest that, depending on their excitation, the HDO lines could be emitted in different regions inside the hot corino where disks could be present. Observations at very high spatial resolution, such as with the Atacama Large Millimeter/Submillimeter Array (ALMA), as well as more complex modeling would then be required to answer this question. Another simpler explanation could arise from the H$_2$ density assumed in the hot corino. If the density used by \citet{Taquet2013} were slightly higher, it could lead to a lower HDO/H$_2$O ratio, reconciling the results.

Assuming that the presence of HDO and H$_2$O in the warm inner envelope is only due to the desorption from the grains, grain-surface chemical models \citep{Cazaux2011,Taquet2013b} can be used to determine the initial conditions (temperature and H$_2$ density) of the parent cloud leading to the HDO/H$_2$O ratios obtained here. The model by \citet{Cazaux2011}, which studies the formation of deuterated ices by accreting species from the gas phase during a free-fall collapse, shows that low temperatures ($\sim$ 12 K) are required to obtain HDO/H$_2$O ratios between 10$^{-4}$ and a few 10$^{-3}$ (see Fig. 4 in \citealt{Coutens2013}). The density is, however, not constrained. Additional information such as D$_2$O/H$_2$O ratios would be necessary to estimate it. The model by \citet{Taquet2013b}, which follows the multilayer formation of deuterated ices with a pseudo time-dependent approach, confirms that the temperature should be low ($\sim$ 10 K) to obtain HDO/H$_2$O ratios close to 10$^{-3}$. Low densities ($\sim$ 10$^3$ cm$^{-3}$) would also be needed (see Fig. 2 in \citealt{Taquet2013}).

\subsubsection{The cold outer HDO/H$_2$O ratios}
\label{sect_cold_ratio}

\citet{Mottram2013} studied several \textit{Herschel}/HIFI water lines for deriving the H$_2$O abundance profile along the envelope of several low-mass protostars and in particular IRAS~4A.
They showed that, in the colder envelope, an H$_2$O abundance profile with a gradual increase of the abundance with the radius is more consistent with the observations than a drop abundance profile as used here (see Fig. 14 in \citealt{Mottram2013}).
Their best-fit obtained with the drop profile ($X_{\rm out}$ = 3 $\times$ 10$^{-10}$ in the outer envelope and $X_{\rm abs}$ = 3 $\times$ 10$^{-7}$, $N_{\rm abs}$ = 1 $\times$ 10$^{15}$ cm$^{-2}$ in the absorbing layer) can, however, give some ideas on the order of magnitude of cold HDO/H$_2$O ratios.
In IRAS~4A, the HDO/H$_2$O ratio is then estimated at about 4\% in the outer envelope and about 9 $\times$ 10$^{-3}$ in the absorbing layer. The cold HDO/H$_2$O ratios therefore seem to be about a few percent and higher than the warm HDO/H$_2$O ratios.
These results obviously have to be considered carefully, as it is shown that the H$_2$O lines are better fitted with a sloping abundance profile. A proper estimate of the HDO/H$_2$O ratio with similar abundance profiles for HDO and H$_2$O is therefore required to confirm the relatively high values of the cold HDO/H$_2$O ratios. 
These high estimates of the cold HDO/H$_2$O ratios are also in agreement with the cold ratios found in IRAS~16293-2422 by \citet[][0.3--1.5 $\times$ 10$^{-2}$]{Coutens2012} and in NGC~1333 IRAS~2A by \citet[][0.9--18 $\times$ 10$^{-2}$]{Liu2011}.
No analysis of the H$_2$O lines observed towards IRAS~4B was carried out; consequently, the cold HDO/H$_2$O ratio cannot be determined in this source at the moment. 

In IRAS~4A, the cold HDO/H$_2$O ratio seems to be higher than the warm HDO/H$_2$O ratio. Similar conclusions were obtained by \citet{Coutens2013} with the study of the D$_2$O/H$_2$O ratio in IRAS~16293-2422.
Different mechanisms were proposed to explain this difference. The first is that water could be formed differently in the warm and cold regions. We can imagine, for example, that, in addition to the desorption of the water molecules trapped in the icy grain mantles (thermal desorption in the hot corino and photodesorption in the outer envelope), water could also be formed in the gas phase through ion-molecule reactions in the cold outer envelope, giving a HDO/H$_2$O abundance ratio up to 1 \citep{Roberts2004}, and through neutral-neutral endothermic reactions in the warm regions, giving a HDO/H$_2$O abundance ratio of 10$^{-3}$--10$^{-2}$ \citep{Thi2010}. As the D/H ratio would lead to a higher deuterium fractionation for cold environments, it could explain why the HDO/H$_2$O ratio is higher in the cold outer envelope. Self-shielding in the hot corino was also suggested in \citet{Coutens2013}. Since H$_2$O is more abundant than its deuterated forms, it should self-shield first, reducing its photodissociation comparatively to HDO and D$_2$O.
Another hypothesis is provided by chemical models including a multi-layer approach. Interstellar ices display a gradient of their deuterium fractionation, the D/H ratio increasing towards the surface of the ices. External ice layers are enriched in deuterium because they have been formed in prestellar cores where the gaseous D/H ratio is high, while the internal part of the ices have been formed in molecular cloud conditions with limited deuterium fractionation. The first outer layers, which are highly deuterated, are then evaporated in the outer envelope through non-thermal processes (photo-evaporation or reactive evaporation), whereas the less deuterated inner layers are only released in the gas phase in the warm envelope when the temperature reaches $\sim$ 100 K (Taquet et al. in prep.). The difference in the HDO/H$_2$O ratio could then originate from the grain layer-structure.

\section{Conclusions}

Deuterated water was detected in the low-mass protostars NGC~1333 IRAS~4A and IRAS~4B with the \textit{Herschel}/HIFI instrument, as well as with ground-based telescopes (IRAM, JCMT, and APEX). The HDO fundamental 1$_{1,1}$--0$_{0,0}$ and 1$_{0,1}$--0$_{0,0}$ transitions observed at 894 and 465 GHz, respectively, show a broad component tracing the outflows in addition to an inverse P-Cygni profile indicating infall motions in the protostellar envelope. 

A RATRAN non-LTE model was then carried out to determine the HDO abundances in the protostellar envelope, after subtraction of the broad component observed on the line profiles.
In IRAS~4A, the derived HDO abundances are about 7.5 $\times$ 10$^{-9}$ in the warm inner regions of the envelope (T $>$ 100 K) where water molecules thermally desorb from the grain mantles, and about 1.2 $\times$ 10$^{-11}$ in the outer envelope. In IRAS~4B, the inner abundance is about 1.0 $\times$ 10$^{-8}$, whereas the outer abundance is about 1.2 $\times$ 10$^{-10}$.
Using the H$_2$$^{18}$O column density determined thanks to the interferometric observations of the 3$_{1,3}$--2$_{2,0}$ transition, we obtain in the warm inner regions HDO/H$_2$O ratios of about 4 $\times$ 10$^{-4}$ -- 3.0 $\times$ 10$^{-3}$ for IRAS~4A and 1 $\times$ 10$^{-4}$ -- 3.7 $\times$ 10$^{-3}$ for IRAS~4B. Comparing the HDO outer abundance with the H$_2$O abundance estimated by \citet{Mottram2013} in IRAS~4A, the outer HDO/H$_2$O ratio seems to be much higher (about a few percent) than in the hot corino. Several mechanisms were suggested in Sect. \ref{sect_cold_ratio} to explain this variation. 

The presence of an extended absorbing layer with a HDO column density of $\sim$ 1.4 $\times$ 10$^{13}$ cm$^{-2}$ is required to reproduce the deep absorbing components seen in the fundamental lines of both sources.
Such a layer was already discovered around the low-mass protostar IRAS~16293-2422 \citep{Coutens2012}. This layer could then be ubiquitous in the surroundings of low-mass protostars.
A narrow SiO emission line detected in the IRAS4 complex by \citet{Lefloch1998} appears at the same velocity as the HDO absorbing components and shows a similar linewidth, which could suggest a common origin of these two species (see Appendix \ref{sect_sio}). Photodesorption mechanisms or sputtering due to decelerated shocks could release into the gas phase molecules of water and SiO contained in the grain mantles.

In the outflows, the HDO column density is estimated, using the RADEX non-LTE code, at $\sim$ 2 -- 4 $\times$ 10$^{13}$ cm$^{-2}$, which leads to an abundance of about 0.7 -- 1.9 $\times$ 10$^{-9}$. The HDO/H$_2$O ratio is estimated at $\sim$ 1 $\times$ 10$^{-3}$ -- 9 $\times$ 10$^{-2}$ in the red part of the outflow and 7 $\times$ 10$^{-4}$ -- 6 $\times$ 10$^{-2}$ in the blue part. These results are consistent with the water deuterium fractionation in the warm inner regions of IRAS~4A derived here, and with the results of \citet{Taquet2013}. This could mean that water is released from the grain mantles by sputtering mechanisms in the outflows and by thermal desorption in the hot corino, supporting an early formation of H$_2$O and HDO during the star formation. However, the wide range of values determined in the outflows does not allow us to clearly assert it.

Finally, we also compared the HDO abundances derived in low-mass protostars.
The outer abundances are particularly well constrained by the HDO 1$_{1,1}$--0$_{0,0}$ line observed at 894 GHz with \textit{Herschel}/HIFI.
The source NGC~1333 IRAS~4A shows the lowest HDO abundances among the low-mass protostars. But the range of HDO outer abundances is relatively narrow, between 10$^{-11}$ and a few 10$^{-10}$. 
A correlation is observed between the ratio of the inner and outer abundances and the submillimeter luminosity, but more observations on a larger sample are required to confirm this correlation. 
For the same region, the HDO outer abundances also seem to vary with the submillimeter luminosity, which could reflect a more efficient non-thermal desorption in less dense envelopes.

\begin{acknowledgements}
  The authors thank D. C. Lis for his comments on the manuscript.
   HIFI has been designed and built by a consortium of institutes and
  university departments from across Europe, Canada, and the United
  States under the leadership of SRON Netherlands Institute for Space
  Research, Groningen, The Netherlands, and with major contributions
  from Germany, France, and the US. Consortium members are: Canada:
  CSA, U.Waterloo; France: IRAP (formerly CESR), LAB, LERMA, IRAM; Germany: KOSMA,
  MPIfR, MPS; Ireland, NUI Maynooth; Italy: ASI, IFSI-INAF,
  Osservatorio Astrofisico di Arcetri-INAF; Netherlands: SRON, TUD;
  Poland: CAMK, CBK; Spain: Observatorio Astron\'omico Nacional (IGN),
  Centro de Astrobiolog\'{\i}a (CSIC-INTA). Sweden: Chalmers
  University of Technology - MC2, RSS \& GARD; Onsala Space
  Observatory; Swedish National Space Board, Stockholm University -
  Stockholm Observatory; Switzerland: ETH Zurich, FHNW; USA: Caltech,
  JPL, NHSC.  
  
  %AC and CV thank the CNES (Centre National d'Etudes Spatiales) for its financial support.
 \end{acknowledgements}

\bibliographystyle{aa}
\bibliography{biblio_NGC1333_hdo}

\newpage

\appendix

\section{\textit{Herschel}/HIFI observations}

\begin{table}[!h]
\renewcommand{\arraystretch}{1.2}
\caption{List of the \textit{Herschel}/HIFI obsIDs}
\begin{tabular}{c c c}
\hline \hline
Source & HDO 1$_{1,1}$--0$_{0,0}$ & HDO 2$_{1,1}$--2$_{0,2}$  \\
\hline
NGC 1333 IRAS~4A & 1342225938 & 1342225931 \\
NGC 1333 IRAS~4B & 1342225940 & 1342225933 \\
Red outflow IRAS~4A & 1342225939 & 1342225932 \\
\hline
\end{tabular}
\label{obs_ID}
\end{table}%

\begin{figure}[!ht]
\begin{center}
\includegraphics[scale=0.83]{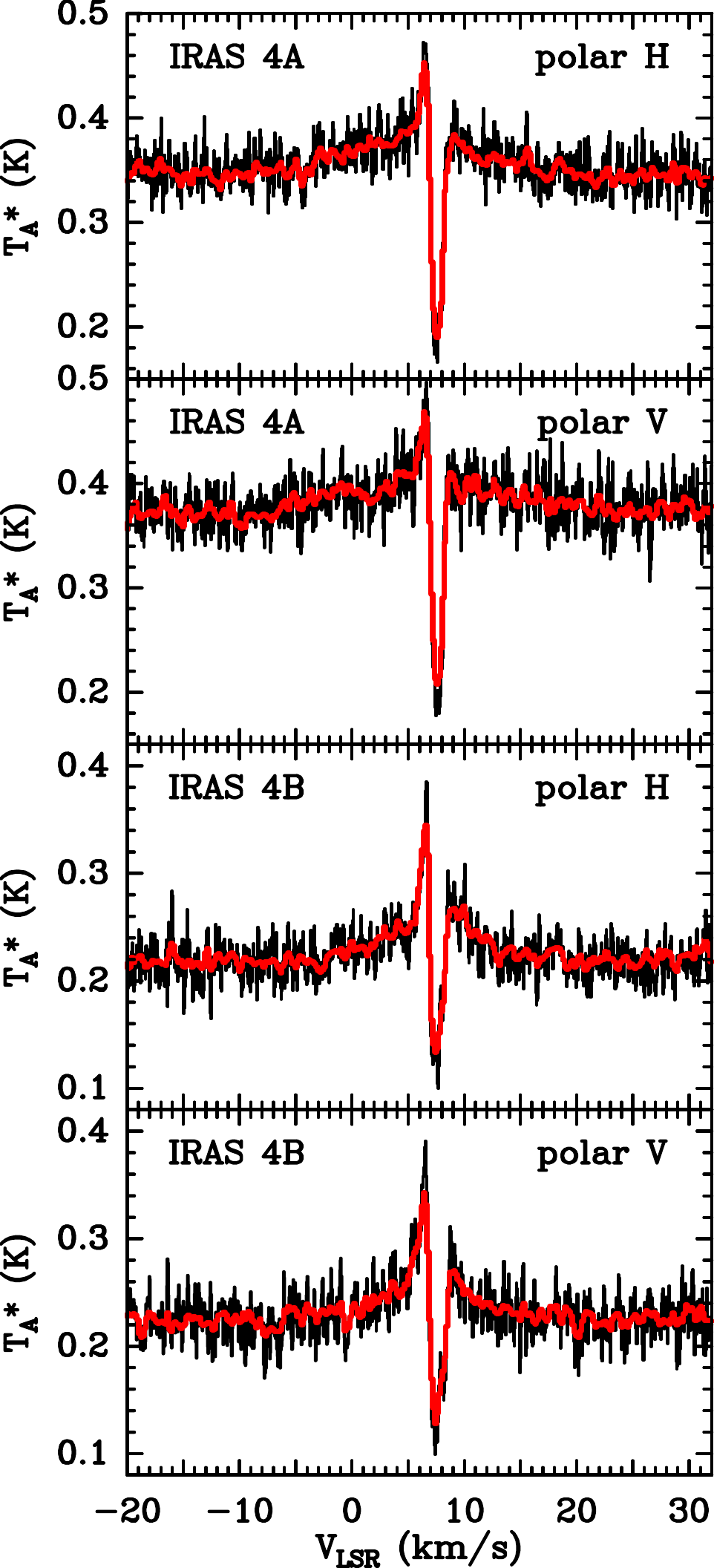}
\caption{Comparison of the WBS (red) and HRS (black) spectra obtained at 894 GHz with \textit{Herschel}/HIFI towards IRAS~4A and IRAS~4B for the H and V polarizations. }
\label{hrs}
\end{center}
\end{figure}

\section{SiO and HDO: The same origin?}
\label{sect_sio}

\begin{figure}[!t]
\begin{center}
\includegraphics[scale=0.44]{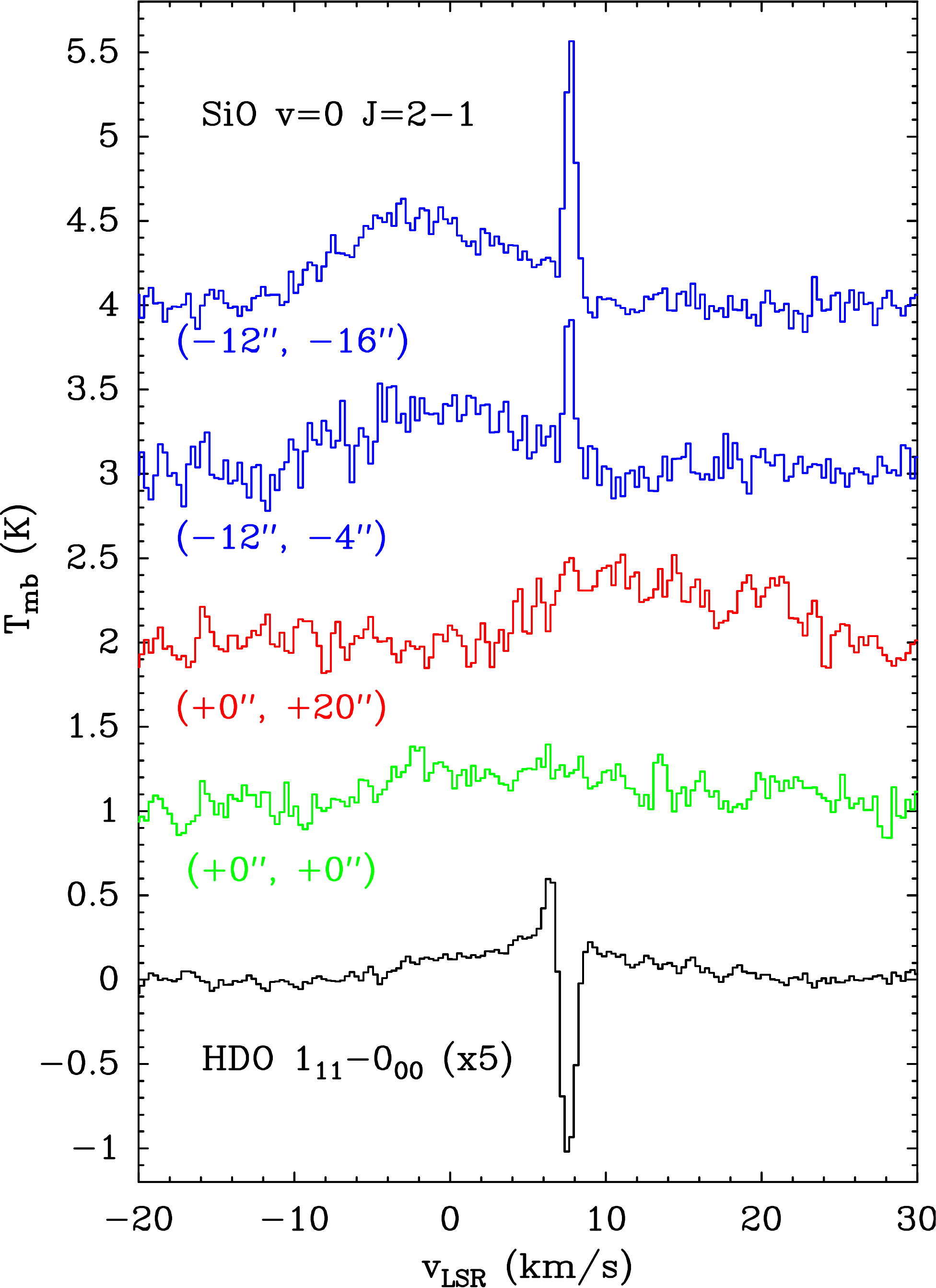}
\caption{Comparison of the HDO 1$_{1,1}$--0$_{0,0}$ transition (\textit{black}) observed towards IRAS~4A and the SiO $\varv$=0 $J$=2--1 line observed towards IRAS~4A (\textit{green}), a position in the red part of the outflow (\textit{red}, position ($+0\arcsec$, $+20\arcsec$) with respect to IRAS4A) and two positions in the blue part of the outflow (\textit{blue}, upper spectra: position ($-12\arcsec$, $-16\arcsec$), lower spectra: position ($-12\arcsec$, $-4\arcsec$)). The narrow emission line of SiO appears at the same velocity as the thin absorbing layer of HDO.}
\label{sio_comp}
\end{center}
\end{figure}

The present observations towards the IRAS~4 region reveal an interesting similarity between the deep absorbing components of HDO and the narrow SiO emission line detected by \citet{Lefloch1998}.
The latter shows a velocity and a linewidth consistent with the HDO absorbing component (see Fig. \ref{sio_comp}), which could suggest a common origin. 
The narrow SiO emission is quite extended ($\sim$ 200$\arcsec$, see Fig. 2 in \citealt{Lefloch1998}), as we have also inferred for the HDO absorbing layer (see Sect. \ref{abs_layer}). 
Variations of the narrow SiO line intensity are observed along the NGC~1333 complex, with particularly bright lines around the blue lobe and non-detections towards the source and the red lobe. From LVG (Large Velocity Gradient) modeling of the bright SiO narrow components in the blue lobe, \citet{Lefloch1998} estimated the SiO column density of 1--5 $\times$ 10$^{12}$ cm$^{-2}$ and the H$_2$ density at about 1--5 $\times$ 10$^5$ cm$^{-3}$, assuming a gas temperature of 33 K equal to that of the dust. 
For the same SiO column density and a lower H$_2$ density ($\sim$10$^4$ cm$^{-3}$), the intensity of the SiO $\varv$=0 $J$=2--1 line is only 0.07 K, in agreement with the non-detection by Lefloch et al. of narrow SiO emission towards our HDO spectra positions ($\lesssim$ 0.27 K for both the source position and the red lobe).
To reproduce the HDO absorbing components with the RATRAN modeling, it is also necessary to restrain the H$_2$ density at a lower value ($< 10^5$ cm$^{-3}$; see Sect. \ref{abs_layer}). A layer of constant column density but subject to density variations ($n$(H$_2$) $<$ 10$^5$ cm$^{-3}$ towards the sources and the red lobe and $n$(H$_2$) $\sim$ 1--5 $\times$ 10$^5$ cm$^{-3}$ towards the blue lobe) could then explain both the deep HDO absorptions and the lack of 
detectable SiO narrow emission towards the sources and the red lobe as well as the bright SiO narrow emission towards the blue lobe.
Alternatively, a layer of uniform low density $\sim$10$^4$ cm$^{-3}$, but with gas temperature locally increasing to $\sim$80--100~K towards the blue lobe, would explain why SiO is only detected in this region. 
In these cases, the HDO absorptions should be shallower towards the blue part of the outflow. Unfortunately, no HDO observation is available at this position to confirm it. 
Assuming that the HDO column density of the absorbing layer does not vary significantly between IRAS 4A and the blue outflow lobe, the SiO/HDO abundance ratio in this layer would be $\sim$ 0.07--0.35.
The beam sizes of the HDO 1$_{1,1}$--0$_{0,0}$ line at 894 GHz and the SiO $\varv$=0 $J$=2--1 transition mapped by \citet{Lefloch1998} with the IRAM-30m telescope are similar.

Another argument suggesting that the narrow HDO and SiO components may trace the same layer is that, in dense and dark clouds, both of these species are believed to be mostly trapped in dust grains and their associated icy mantles. Hence, the same mechanism could release SiO (or other Si components) at the same time as HDO and H$_2$O into the gas phase.
We discuss below two mechanisms able to release these molecules into the gas phase: photodesorption and shocks.
Another mechanism proposed by \citet{Yildiz2012} to explain detections of narrow $^{13}$CO emission lines within the outflow lobes of IRAS~4A is UV-heating by the central source. The narrow SiO emission in NGC~1333 has a more extended spatial distribution, which instead tends to peak outside of the outflow cavities (see Fig. 1 in \citealt{Lefloch1998}), which appears to rule out this interpretation here.

\subsection{Photodesorption}

It was suggested that narrow SiO emission at the systemic velocity could be produced by photodesorption mechanisms in photodissociation regions (PDR) such as the Orion bar \citep{Walmsley1999, Schilke2001}. Indeed, the strength of Si$^+$ in irradiated regions suggests that up to $\sim$10\% of the elemental silicon is not locked in silicate grain cores, but in less refractory material that gets efficiently desorbed at low $A_V$ \citep{Haas1986}. 
We note that a similar fraction of silicon ($\sim$ 5\%) with respect to the cosmic abundance ($\sim$ 3 $\times$ 10$^{-5}$) is measured in the gas-phase of diffuse clouds \citep[e.g.,][]{Morton1975,Sofia1994,Miller2007}.
\citet{Gusdorf2008} also estimated that, to account for the observed SiO line intensities in the blue lobe of L1157, about $\sim$10\% of the silicon assumed in the form of SiO has to be locked in the grain mantles. 

Photodesorption could then release SiO in the gas phase simultaneously with water and HDO. 
Alternatively, the  material could be trapped in grain mantles in the form of atomic Si; the Si would then be desorbed and react with O$_2$ or OH to form SiO according to the following reactions \citep{Walmsley1999}:
\begin{equation}
\rm Si +O_2 \rightarrow SiO + O 
\end{equation}
\begin{equation}
\rm Si +OH \rightarrow SiO + H.
\end{equation}
The balance between photodesorption and photodissociation predicts an SiO-enriched layer located at some depth behind the photodissociation front, which can easily explain the SiO column density observed towards the Orion Bar of a few $10^{12}$ cm$^{-2}$ \citep{Walmsley1999, Schilke2001}. Although O$_2$ is actually less abundant than assumed in these early models, a tentative detection of O$_2$ was recently reported towards IRAS 4A at a velocity ($\sim$ 8.0 $\kms$) close to our HDO absorptions ($\sim$ 7.6 $\kms$), and suggested to be produced by photodesorption mechanisms \citep{Yildiz2013}.
Photodissociation of H$_2$O would also produce ample OH to react with Si and form SiO by the second reaction above. 

Photodesorption of grain mantles thus appears as a possible explanation for the narrow HDO and SiO components at the systemic velocity towards the IRAS 4A region. Photodissociation region models appropriate for the surface of the NGC1333 cloud, i.e., with lower UV radiation field ($\chi$ $\sim$ 1) than the Orion bar ($\chi$ $\sim$ 10$^4$--10$^5$), could verify if this mechanism also reproduces the observed SiO/HDO ratio.

\subsection{Shocks}

Shocks are another non-thermal mechanism able to remove both silicon and water from dust grains. Contrary to CO that traces all of the entrained material in molecular flows, broad SiO emission is mostly detected in the youngest outflows from Class 0 protostars \citep{Gibb2004}. This is not surprising, as SiO requires a higher density than CO for excitation and especially shocks fast enough ($\ge 10-25$ \kms) to efficiently sputter SiO and Si from the mantles and grain cores in which they would otherwise remain locked \citep{Schilke1997,Caselli1997, Gusdorf2008a,Gusdorf2008,Guillet2009}. Sputtering processes also efficiently release water ice from grain mantles, as illustrated, for example, by the similarity of the SiO and H$_2$O high-velocity wings in the L1157 outflow \citep{Lefloch2010}, and the broad wings in the HDO profile towards IRAS~4A (see Fig.~\ref{sio_comp}). The IRAS~4A outflow wings
have a total SiO column density of a few $10^{13}$ cm$^{-2}$ and an abundance relative to CO $\sim$ 3 $\times$ $10^{-4}$ \citep{Lefloch1998}, typical of young outflows \citep{Tafalla2010}. The narrow SiO component towards IRAS~4A has a  column density that is 10 times smaller than the column density in the broad SiO component, and an abundance relative to CO that is 100 times lower, when compared to the total C$^{18}$O at systemic velocity \citep{Lefloch1998}. Hence, the presence of the narrow SiO and HDO components could be explained in this context if about 10\% of the SiO (and HDO) released in outflow shocks suffered strong deceleration and chemical dilution (by a typical factor of 10--100) due to turbulent mixing with static ambient gas \citep{Lefloch1998,Codella1999}. 
The surface dilution induced by the mixing process would be compensated by the contribution of neighboring outflows, which cover a large fraction of the surface area in NGC 1333.
The SiO/HDO ratio is estimated in the outflows at 0.25--1.5 (using $N$(SiO) = 1--3 $\times$ 10$^{13}$ cm$^{-2}$ from \citealt{Lefloch1998} and $N$(HDO) = 2--4 $\times$ 10$^{13}$ cm$^{-2}$ from Sect. \ref{sect_outflow}). This could be consistent with that estimated for the narrow layer (0.07--0.35), raising the possibility that shocked gas could contribute to this extended SiO/HDO layer. 
The SiO and HDO molecules will then re-deplete onto grain mantles on timescales depending on the volume densities ($\sim$ 10$^4$ years for densities about 10$^5$ cm$^{-3}$ and $\sim$ 10$^5$ years for densities about 10$^4$ cm$^{-3}$).
Numerical simulations would be helpful to compare the predicted SiO/HDO ratio with the observations as well as to verify that sufficient deceleration can indeed be reached before SiO and HDO are readsorbed onto the grains to produce narrow SiO and HDO features close to systemic velocity.

Interactions of magnetic and/or radiative shock precursors with the ambient pre-shocked clumpy medium were also mentioned as a potential mechanism leading to the emission of narrow SiO lines \citep{Jimenez2004,Jimenez2010,Jimenez2011}. This interpretation was supported by the higher degree of excitation of the ion fluid compared to the neutral fluid towards the protostar L1448-mm.
\citet{Roberts2012} showed, however, that the higher excitation of H$^{13}$CO$^+$ in the narrow component is not a conclusive indication of a precursor (it could simply be due to protostellar heating).
This interpretation then relies entirely on the fact that the narrow SiO emission is compact and spatially confined to the regions around protostars. In the NGC~1333 complex, the narrow SiO component is spatially extended over the whole region \citep{Lefloch1998}. Therefore, the shock precursor interpretation would not hold here.

\end{document}